\documentclass[english,aps,numerical,nofootinbib]{revtex4-1}
\usepackage[T1]{fontenc}
\usepackage[latin9]{inputenc}
\usepackage{amsmath}
\usepackage{graphicx}
\usepackage{amssymb}
\usepackage{epsfig}
\usepackage{marginnote}

\makeatletter

\providecommand{\tabularnewline}{\\}

\newcommand{\lsim}{\mbox{\raisebox{-.9ex}{~$\stackrel{\mbox{$<$}}{\sim}$~}}}

\makeatother

\usepackage{babel}

\makeatother

\usepackage{babel}

\begin{document}
\global\long\global\long\def\s#1{s_{#1}}
 \global\long\global\long\def\Cpi{C_{1}^{+}}
 \global\long\global\long\def\Cpii{C_{2}^{+}}
 \global\long\global\long\def\Cmi{C_{1}^{-}}
 \global\long\global\long\def\Cmii{C_{2}^{-}}

\global\long\global\long\def\t#1{\tilde{#1}}
 \global\long\global\long\def\re{\mathrm{e}}
 \global\long\global\long\def\e#1{#1_{\re}}
 \global\long\global\long\def\x#1{#1_{\mathrm{x}}}
 \global\long\global\long\def\xx#1#2{#1_{#2\mathrm{x}}}
 \global\long\global\long\def\y#1{#1_{y}}
 \global\long\global\long\def\b#1{\bar{#1}}

\global\long\global\long\def\tx#1{\x{\t{#1}}}
 \global\long\global\long\def\txx#1#2{\xx{\t{#1}}{#2}}
 \global\long\global\long\def\te#1{\e{\t{#1}}}
 \global\long\global\long\def\ty#1{\y{\t{#1}}}

\global\long\global\long\def\Cbp#1{\b C_{#1}^{+}}
 \global\long\global\long\def\Cbm#1{\b C_{#1}^{-}}

\global\long\global\long\def\mpl{m_{\mathrm{Pl}}}
 \global\long\global\long\def\anm{Q}
 \global\long\global\long\def\ow{\Omega_{\mathrm{W}}}

\global\long\global\long\def\fnl{f_{\mathrm{NL}}}
 \global\long\global\long\def\fnle{\fnl^{\mathrm{eql}}}
 \global\long\global\long\def\fnls{\fnl^{\mathrm{sqz}}}
 \global\long\global\long\def\fnlf{\fnl^{\mathrm{flt}}}

\title{Parity Violating Statistical Anisotropy}

\author{Konstantinos Dimopoulos}


\affiliation{Consortium for Fundamental Physics, Physics Department, Lancaster
University, Lancaster LA1 4YB, U.K.}

\email{k.dimopoulos1@lancaster.ac.uk}

\author{Mindaugas Kar\v{c}iauskas}

\affiliation{CAFPE and Departamento de F\'isica Te\'orica y del Cosmos, Universidad
de Granada, Granada-18071, Spain}

\email{mindaugas@ugr.es}

\begin{abstract}
Particle production of an Abelian vector boson field with an axial coupling is 
investigated. The conditions for the generation of scale invariant spectra for 
the vector field transverse components are obtained. If the vector field 
contributes to the curvature perturbation in the Universe, scale-invariant 
particle production enables it to give rise to statistical anisotropy in the 
spectrum and bispectrum of cosmological perturbations. The axial coupling 
allows particle production to be parity violating, which in turn can generate 
parity violating signatures in the bispectrum. The conditions for parity 
violation are derived and the observational signatures are obtained in the 
context of the vector curvaton paradigm. Two concrete examples are presented 
based on realistic particle theory.
\end{abstract}

\maketitle

\section{Introduction}

In the last few years the contribution of cosmic vector fields to the curvature
perturbation in the Universe is under investigation. This effort was triggered 
by the pioneering work in Ref.~\cite{vecurv}, which introduced the vector 
curvaton mechanism, through which vector boson fields can contribute to or even
fully generate the curvature perturbation (for a recent review see 
Ref.~\cite{vcrev}). In analogy with the scalar curvaton 
mechanism \cite{curv}, the vector curvaton is a spectator field during cosmic 
inflation, which becomes heavy after the end of inflation, when it can dominate
(or nearly dominate) the density of the Universe before its decay, thereby 
imprinting its contribution to the curvature perturbation $\zeta$. It was soon 
realised that, through their contribution to $\zeta$, vector fields can give 
rise to distinct observational signatures, namely produce statistical 
anisotropy in the spectrum and bispectrum of $\zeta$ 
\cite{yokosoda,stanis,fnlanis}. This is a new observable, which amounts to 
direction dependent patterns on the CMB temperature anisotropies, which can be
predominantly anisotropic in the bispectrum \cite{fnlanis,bartolo}. Such 
patterns cannot be generated if scalar fields alone are responsible for the 
formation of $\zeta$. Thus, in view of the imminent observations of the Planck 
satellite mission that may well observe statistical anisotropy, it is of 
paramount importance to investigate this effect and what it can reveal to us 
for the underlying theory. In particular, if statistical anisotropy is observed
by Planck, we will have our first glimpse into the gauge field content of 
theories beyond the standard model.

In general, one can divide the mechanisms that have been developed so far for 
the contribution of vector fields into $\zeta$ in two classes. Firstly, one can
have an {\em indirect} influence of the vector field to inflation. One way to 
do this is by considering that the vector field develops a condensate, which 
generates anisotropic stress that renders inflation mildly anisotropic. In 
this case, the anisotropy in the expansion reflects itself onto the 
perturbations of the inflaton field, thereby generating statistical anisotropy 
in $\zeta$ \cite{anisinf,attract} (for a recent review see \cite{sodarev}). 
This anisotropisation of inflation is usually achieved by introducing some 
coupling between the vector field and the inflaton \cite{anisinf,attract,sodarev,fF2more}. Alternatively, such 
coupling can backreact to the generation of the inflaton perturbations 
regardless of anisotropising the expansion \cite{pelosononG}. 

The other class of mechanisms which give rise to statistical anisotropy 
considers the {\em direct} contribution of vector field perturbations to 
$\zeta$. One way this can be done is through the vector curvaton mechanism 
mentioned above \cite{vecurv,sugravec,nonmin,stanis,varkin,dbivc,etavec}, 
where perturbations of the vector field perturb its energy density and, hence, 
the moment of (near) domination of the Universe by the vector field condensate.
Statistical anisotropy in this case is due to the fact that vector fields 
undergo anisotropic particle production, in general \cite{stanis,varkin}. Other
mechanisms have been employed as well, such as the end of inflation mechanism
\cite{yokosoda,stanis}. Finally, one way to use directly the perturbations of 
vector fields to source $\zeta$ is by considering a large number of them acting
as inflatons as in Ref.~\cite{vecinf}. 

What do observations say about statistical anisotropy in the curvature 
perturbation? The observed bound on statistical anisotropy in the power 
spectrum of $\zeta$ is surprisingly low; it seems that as much as 30\% of it
is still allowed. In fact, statistical anisotropy at this level was reported 
in Ref.~\cite{9sigma} at the level of 9-$\sigma$! However, the 
direction of the anisotropy was suspiciously close to the ecliptic plane so
the authors of these studies conclude that the finding is probably due to some 
systematic mistake, hence being treated as an upper bound. The Planck mission 
will reduce this bound down to 2\% if statistical anisotropy is not observed 
\cite{kamion}. This means that the so-called anisotropy parameter, which 
quantifies statistical anisotropy in the spectrum of $\zeta$, must lie in 
the following range if it is to be observed in the near future:
\begin{equation}
0.02\lesssim g_\zeta\lesssim 0.3\,.
\label{grange}
\end{equation}

Although, in principle, statistical anisotropy in $\zeta$ can be scale 
dependent, a scale invariant spectrum of vector field perturbations results in
scale-independent $g_\zeta$, which does not have to be fine-tuned such that it 
falls into the above range on the cosmological scales. Moreover, even if 
Eq.~(\ref{grange}) is satisfied for cosmological scales, a strongly tilted 
spectrum would generate intense anisotropy, giving rise to excessive curvature 
perturbations (e.g. leading to copious primordial black hole formation) or 
destabilising inflation itself. Thus, a (nearly) scale-invariant spectrum of 
vector field perturbations is preferred. As shown in Ref.~\cite{attract}, this 
can be naturally attained in certain types of theories.

Statistical anisotropy generates angular modulation of the power spectrum, but also of higher order correlators. Moreover, even if the anisotropy in the spectrum satisfies the upper bound in Eq.~\eqref{grange}, higher order correlators can be predominantly anisotropic (see e.g. \cite{varkin}). It is thus important to study the effects of statistical anisotropy on higher order correlators \cite{anisoCorrs} as well as develop methods of detecting these effects in the CMB temperature perturbation \cite{anisofNL}.

Not many models exist as yet, for the formation of a superhorizon spectrum of 
vector field perturbations during inflation. The problem is rather old as it 
was investigated, at first, in order to generate a primordial magnetic field
during inflation, with superhorizon coherence. A massless Abelian vector boson 
field, such as the photon, cannot undergo particle production during inflation
because it is conformally invariant. A breakdown of its conformality is, 
therefore, required in order to form the desired superhorizon spectrum of 
perturbations. Modifications of the theory in order to attain such breakdown 
were originally investigated in Ref.~\cite{TW}, by coupling electromagnetism 
non-minimally to gravity. Some of these proposals were recently implemented in 
the effort to generate a contribution of vector field to $\zeta$; notably a 
non-minimal coupling to gravity of the form $RA_\mu A^\mu$. Such coupling  was 
employed in Ref.~\cite{vecinf} where hundreds of vector fields are used as 
inflations, and also in Refs.~\cite{nonmin,stanis}, in the context of the 
vector curvaton model. However, this proposal was criticised for giving rise to
ghosts \cite{ghosts} (see however Ref.~\cite{RA2save}).

Another model, which does not suffer from instabilities, is considering the 
supergravity-inspired varying kinetic function for the vector field 
$f(t)F_{\mu\nu}F^{\mu\nu}$. The latter has been shown to give rise to a new 
inflationary attractor under fairly general 
conditions \cite{anisinf,attract}, which leads to scale-invariant vector field 
perturbations \cite{attract}, when the kinetic function is modulated by the 
inflaton field. This model also has a long history, since it has been used to
generate primordial magnetic fields too \cite{gaugekin}. More recently, 
however, it has been used to affect $\zeta$ in the vector curvaton mechanism
\cite{stanis,varkin,dbivc,sugravec,etavec} as well as the end of inflation 
mechanism \cite{yokosoda,EoInonAbel}. 

A third popular choice for primordial magnetic field generation during 
inflation is considering the axial term $h(t) F_{\mu\nu}\tilde F^{\mu\nu}$
\cite{axialpmf}. Recently, some works have investigated the vector field
backreaction onto inflation, if the axial coupling $h$ is modulated by a
pseudo-scalar inflaton field. Ref.~\cite{sorbo} suggests that the axial term, 
can allow steep inflation with sub-Planckian axion decay constant, evading 
thereby the basic problem of natural inflation \cite{natural}. In 
Ref.~\cite{pelosononG}, the backreaction of the generated vector field 
perturbations onto the inflaton perturbations is shown to generate significant 
non-Gaussianity in the latter. Finally, the effects of the axial coupling onto
gravitational waves is investigated in Ref.~\cite{axialgw}, which may provide 
parity violating signatures on the tensor modes and their effects onto the~CMB.\footnote{Other works on parity violation in the graviton bispectrum can be found in Ref.~\cite{prVlGr}.}

Parity violation is a special property of the axial model, compared to the
other two mentioned above ($fF^2$ and $RA^2$), which are parity conserving. 
What would it amount of if the vector field perturbations {\em directly} 
contributed to $\zeta$, e.g. through the vector curvaton mechanism? 
As shown in Ref.~\cite{stanis} parity violation does not feature in the power 
spectrum of the density perturbations. However, in Ref.~\cite{fnlanis} it was 
found that parity violation does affect the bispectrum of $\zeta$ and it should
reflect itself onto the shape and amplitude of the anisotropic $f_{\rm NL}$. 
In this paper we present a complete study of the particle production of an 
Abelian vector boson field featuring both a varying kinetic function and a 
non-trivial axial coupling.\footnote{The contribution of non-Abelian vector 
fields to $\zeta$ has been considered in Refs.~\cite{EoInonAbel,nonAbel}.}  We obtain scale
invariant parity violating spectra and we single out the conditions for their 
successful generation, providing some concrete examples based on particle 
theory. Finally, we apply our findings onto the vector curvaton paradigm and 
find the generated $g_\zeta$ and $f_{\rm NL}$, which are to be contrasted with 
observations.

The structure of our paper is as follows. In Sec.~\ref{sec:anisotropic-fNl}
we present a brief review of the general case when a vector field contributes 
directly into the spectrum and bispectrum of $\zeta$. In Sec.~\ref{sec:pvvf}
we present the axial model and investigate particle production throughout the 
parameter space. In Sec.~\ref{sub:Scale-Invariant} we focus on the most 
promising case which produces scale-invariant parity violating spectra for the 
vector field components and apply our findings to the vector curvaton 
mechanism. In Sec.~\ref{ce} we discuss two concrete examples based on realistic
particle theory. We conclude in Sec.~\ref{conc}.
Throughout our paper we consider natural units where \mbox{$c=\hbar=k_B=1$} and
Newton's gravitational constant is \mbox{$8\pi G=m_P^{-2}$}, with 
\mbox{$m_P=2.4\times 10^{18}\,$GeV} being the reduced Planck mass.
We use the metric $g_{\mu\nu}=\mathrm{diag}\left(1,-a^{2},-a^{2},-a^{2}\right)$
and assume (quasi) de Sitter inflation with Hubble parameter 
\mbox{$H\approx\,$constant}.

\section{The Anisotropic, Parity Violating $\fnl$\label{sec:anisotropic-fNl}}

In this section we briefly summarize the results in 
Refs.~\cite{stanis,fnlanis}
and calculate the anisotropic $\fnl$ in the so called {}``flattened''
shape for the first time.

Massive vector fields have three degrees of freedom. During inflation,
after horizon exit quantum fluctuations of these degrees of freedom
become classical perturbations. To deal with these perturbations it
is convenient to use circular polarization vectors. As these vectors
transform differently under rotations, Lorentz invariance of the Lagrangian
guarantees that equations for the three polarizations are uncoupled.
Let us denote the power spectrum of left- and right-handed polarization
modes by $\mathcal{P}_{\mathrm{L}}$ and $\mathcal{P}_{\mathrm{R}}$
respectively and the longitudinal one by $\mathcal{P}_{\|}$. Furthermore,
expressions for the power spectrum and higher order correlators of
the primordial curvature perturbation $\zeta$ become much simpler
expressed in terms of $\mathcal{P}_{\pm}$, which are defined as
\begin{equation}
\mathcal{P}_{\pm}=\frac{1}{2}\left(\mathcal{P}_{\mathrm{R}}\pm\mathcal{P}_{\mathrm{L}}\right),
\end{equation}
 where $\mathcal{P}_{+}$ correspond to parity conserving and $\mathcal{P}_{-}$
parity violating spectrum. The later is non-zero only if the left-
and right- handed polarizations acquire different perturbation spectrum.
Calculating $\fnl$ we will also find convenient to normalize spectra
as
\begin{equation}
p\equiv\frac{\mathcal{P}_{\|}-\mathcal{P}_{+}}{\mathcal{P}_{+}}\quad\mathrm{and}\quad q\equiv\frac{\mathcal{P}_{-}}{\mathcal{P}_{+}}.\label{eq:pq-def}
\end{equation}
If the $p$ and/or $q$  parameters are non-zero the vector field perturbation is statistically anisotropic. Note that by definition the $p$ parameter can take values $p \geq -1$ and the parity violation one $-1 \leq q \leq 1$.

To calculate the curvature perturbation $\zeta$ we use the so called
$\delta N$ formalism. This formalism was generalized to include the
perturbation from the vector field in 
Ref.~\cite{stanis}.
Up to the second order it reads
\begin{equation}
\zeta=N_{\phi}\delta\phi+N_{i}^{W}\delta W_{i}+N_{ij}^{W}\delta W_{i}\delta W_{j},\label{eq:dN-formula}
\end{equation}
where $N_{\phi}\equiv\partial N/\partial\phi$, $N_{i}^{W}\equiv\partial N/\partial W_{i}$
and $N_{ij}^{W}\equiv\partial^{2}N/\partial W_{i}\partial W_{j}$
and derivatives are with respect to the homogeneous values of fields.
Also, the summation over repeated spatial indices is assumed. In this
equation we also assumed that the scalar field perturbation is Gaussian,
hence no second order terms in $\delta\phi$, and that scalar and
vector field perturbations are uncoupled.

Using the $\delta N$ formula we can easily find the spectrum of $\zeta$
at tree level \cite{stanis}\footnote{The quadrupole modulation of $\mathcal{P}_{\zeta}$ can also be generated during weakly anisotropic inflationary expansion  \cite{ack} or inflationary models in non-commutative space-times \cite{nonComm}.}
\begin{equation}
\mathcal{P}_{\zeta}\left(\mathbf{k}\right)=\mathcal{P}_{\zeta}^{\mathrm{iso}}\left(k\right)\left[1+g_{\zeta}\left(\hat{\mathbf{k}}\cdot\hat{\mathbf{N}}^{W}\right)^{2}\right],\label{eq:Pz-def}
\end{equation}
 where $k\equiv\left|\mathbf{k}\right|$, $\hat{\mathbf{k}}=\mathbf{k}/k$
and $\hat{\mathbf{N}}^{W}=\mathbf{N}^{W}/N_{W}$ with $N_{W}\equiv\left|\mathbf{N}^{W}\right|$.
In this equation the isotropic part of the spectrum is 
\begin{equation}
\mathcal{P}_{\zeta}^{\mathrm{iso}}\left(k\right)=N_{\phi}^{2}\mathcal{P}_{\phi}\left(k\right)\left(1+\xi\right),
\end{equation}
 where %
\footnote{Note that this definition of $\xi$ reduces to $\beta$ in 
Ref.~\cite{fnlanis}
when $\mathcal{P}_{+}=\mathcal{P}_{\phi}$. In this limit it is also
equal to $\xi$ in 
Refs.~\cite{varkin}.
}
\begin{equation}
\xi\equiv\left(\frac{N_{W}}{N_{\phi}}\right)^{2}\frac{\mathcal{P}_{+}\left(k\right)}{\mathcal{P}_{\phi}\left(k\right)}\label{eq:xsi-def}
\end{equation}
 and it quantifies the contribution of the vector field perturbation
to the total curvature perturbation. For $\xi<1$ the scalar field
contribution dominates.

The amplitude $g_{\zeta}$ of the quadrupole modulation in the spectrum
in Eq.~\eqref{eq:Pz-def} is given by \cite{stanis}
\begin{equation}
g_{\zeta}\left(k\right)=N_{W}^{2}\frac{\mathcal{P}_{\|}\left(k\right)-\mathcal{P}_{+}\left(k\right)}{\mathcal{P}_{\zeta}^{\mathrm{iso}}\left(k\right)}=\frac{\xi}{1+\xi}p\left(k\right).
\label{eq:gz-expr}
\end{equation}
 From the last equation we see that a mildly statistically anisotropic
vector field perturbation, $|p|\ll1$, can generate the total curvature
perturbation, that is $g_{\zeta}\approx p$ with $\xi\gg1$. Then
the observational bound on $g_{\zeta}$, discussed in the Introduction,
gives $p<0.3$. If, on the other hand, $p$ violates this bound, the
vector field can generate only a subdominant contribution to $\zeta$,
i.e. $\xi<1$, in which case $g_{\zeta}\approx\xi\, p$.

Note, that the curvature perturbation power spectrum $\mathcal{P}_{\zeta}$
is proportional only to $p$ but not $q$. Thus the possible parity
violation of the vector field perturbation cannot be detected in the
spectrum of $\zeta$. We need to measure higher order correlators
for that.

In the present work we study the generation of $\zeta$ by the vector
curvaton scenario, in which $N_{i}^{W}$ and $N_{ij}^{W}$ can be
found to be \cite{stanis}
\begin{equation}
N_{i}^{W}=\frac{2}{3}\hat{\Omega}_{W}\frac{W_{i}}{W^{2}}\quad\mathrm{and}\quad N_{ij}^{W}=\frac{2}{3}\hat{\Omega}_{W}\frac{\delta_{ij}}{W^{2}},
\end{equation}
 with $\hat{\Omega}_{W}$ defined as
\begin{equation}
\hat{\Omega}_{W}\equiv\frac{3\rho_{W}}{3\rho_{W}+4\rho_{\mathrm{r}}}=\frac{3\ow}{4-\ow}.\label{eq:hOW-def}
\end{equation}
 In the above $\rho_{W}$ and $\rho_{\mathrm{r}}$ are energy densities
of the vector field and radiation at the curvaton decay and 
\begin{equation}
\ow\equiv\rho_{W}/\rho,\label{eq:Ow-def}
\end{equation}
 where $\rho=\rho_{W}+\rho_{\mathrm{r}}$. If the curvaton decays
while subdominant then $\hat{\Omega}_{W}=3\ow/4<1$.

The non-linearity parameters $\fnl$ for this scenario were calculated
in Ref.~\cite{fnlanis}. 
In this reference two shapes
of $\fnl$ were considered: the equilateral one $\fnle$, with $k_{1}\approx k_{2}\approx k_{3}$,
and the squeezed one $\fnls$, with $k_{1}\approx k_{2}\gg k_{3}$.
For flat perturbation spectra they are given by

\begin{equation}
\frac{6}{5}\fnle=\frac{\xi^{2}}{\left(1+\xi\right)^{2}}\frac{3}{2\hat{\Omega}_{W}}\left[\left(1+\frac{1}{2}q^{2}\right)+\left(p+\frac{1}{8}p^{2}-\frac{1}{4}q^{2}\right)W_{\perp}^{2}\right],\label{eq:fNL-equil}
\end{equation}
 
\begin{equation}
\frac{6}{5}\fnls=\frac{\xi^{2}}{\left(1+\xi\right)^{2}}\frac{3}{2\hat{\Omega}_{W}}\left(1+pW_{\perp}^{2}+ipqW_{\perp}\sqrt{1-W_{\perp}^{2}}\sin\omega\right),\label{eq:fNL-local}
\end{equation}
In these expressions $\mathbf{W}_{\perp}$ is the projection vector
of $\hat{\mathbf{W}}\equiv\mathbf{W}/W$ onto the plane of the three
wave-vectors $\mathbf{k}_{1}$, $\mathbf{k}_{2}$ and $\mathbf{k}_{3}$
with $0\le W_{\perp}\le1$. The angle $\omega$ in the second equation
is between $\mathbf{W}_{\perp}$ and the squeezed $\mathbf{k}$ vector.
Here, we also calculate the third, so called flattened, shape where
the length of one of the wave-vectors is twice as large as of the
other two, e.g. $k_{1}\approx k_{2}\approx\frac{1}{2}k_{3}$ or $\frac{1}{2}k_{1}\approx k_{2}\approx k_{3}$:
\begin{equation}
\frac{6}{5}\fnlf=\frac{\xi^{2}}{\left(1+\xi\right)^{2}}\frac{3}{2\hat{\Omega}_{W}}\left[\left(1-\frac{3}{5}q^{2}\right)+\left(2p+p^{2}+\frac{3}{5}q^{2}\right)\cos^{2}\varphi W_{\perp}^{2}\right].\label{eq:fNL-flat}
\end{equation}
where $\varphi$ is the angle between $\mathbf{W}_{\perp}$ and the longest $\mathbf{k}$ vector.

For the vector field contribution to $\zeta$ with $g_{\zeta}\lsim 0.1$ and $ p \le \mathcal O (1)$ current observation constraints give $\hat \Omega_W > 10^{-2}$, with $|\fnl| \lsim 100$ \cite{komatsu}. The parity violation of the vector field perturbation (non-zero $q$) modulates the shape of $\fnl$ by suppressing its value in the equilateral configuration and enhancing it in the flattened one. It also introduces an imaginary term in the squeezed configuration $\fnls$, which is real in position space by the virtue of the reality condition. From Eqs.~\eqref{eq:fNL-equil}-\eqref{eq:fNL-flat} it is also clear that in general the vector field contribution to $\zeta$ generates an angular modulation of $\fnl$ (terms proportional to $W_\perp$). Large enough statistical anisotropy of the perturbation (parametrized by $|p|$ and $|q|$) can generate predominantly anisotropic $\fnl$ with a configuration dependent amplitude and form of the angular modulation.

\section{Parity Violating Vector Field}\label{sec:pvvf}

\subsection{Equations of Motion \label{sub:EoMs}}

Let us consider the Lagrangian of a massive $U\left(1\right)$ vector
field
\begin{equation}
\mathcal{L}=-\frac{1}{4}fF_{\mu\nu}F^{\mu\nu}-\frac{1}{4}hF_{\mu\nu}\tilde{F}^{\mu\nu}+\frac{1}{2}m^{2}A_{\mu}A^{\mu},\label{eq:Lagrangian}
\end{equation}
 where $F_{\mu\nu}=\partial_{\mu}A_{\nu}-\partial_{\nu}A_{\mu}$ and

\begin{equation}
\tilde{F}^{\mu\nu}=\frac{1}{2}\frac{\epsilon^{\mu\nu\rho\sigma}}{\sqrt{-g}}F_{\rho\sigma},
\end{equation}
 with $\epsilon^{\mu\nu\rho\sigma}$ being the totally antisymmetric
tensor. The three functions $f\left(t\right)$, $h\left(t\right)$
and $m^{2}\left(t\right)$ are time dependent. Their variation is
provided by other dynamical degrees of freedom in the theory. In this
section we do not specify these degrees of freedom. Our aim is rather
to find a scaling of these functions which give a flat perturbation
spectrum for the vector field.

The Lagrangian of the form in Eq.~\eqref{eq:Lagrangian}, with $h=0$
and non-zero mass term, was first proposed in 
Ref.~\cite{sugravec} 
for the generation of the primordial curvature perturbation and then
extensively studied in Refs.~\cite{varkin}.
It was found that during inflation all three degrees of freedom of
the massive vector field acquire a flat perturbation spectrum if the
kinetic function $f$ and the mass $m$ evolves as
\begin{equation}
f\propto a^{-1\pm3}\quad\mathrm{and}\quad m\propto a,\label{eq:f-m-scaling}
\end{equation}
 where $a$ is a scale factor. In this section we study perturbations
of the parity violating vector field. We find the evolution of functions
$f\left(t\right)$, $h\left(t\right)$ and the mass $m\left(t\right)$
which result in a flat spectrum.

The Euler-Lagrange equation with the Lagrangian in Eq.~\eqref{eq:Lagrangian}
gives the field equation for the vector field as
\begin{equation}
\left[\partial_{\mu}+\partial_{\mu}\ln\sqrt{-g}\right]\left(fF^{\mu\nu}+h\tilde{F}^{\mu\nu}\right)+m^{2}A^{\nu}=0.
\end{equation}
 The temporal and spatial components of this equation read as
\begin{equation}
\partial_{i}\dot{A}_{i}-\partial_{i}\partial_{i}A_{0}+\frac{\left(am\right)^{2}}{f}A_{0}=0
\end{equation}
 and 
\begin{equation}
\ddot{A}_{i}+\left(H+\frac{\dot{f}}{f}\right)\dot{A}_{i}+\frac{m^{2}}{f}A_{i}-a^{-2}\left(\partial_{j}\partial_{j}A_{i}-\partial_{i}\partial_{j}A_{j}\right)-a^{-1}\frac{\dot{h}}{f}\epsilon^{ijl}\partial_{j}A_{l}=\partial_{i}\left[\dot{A}_{0}+\left(H+\frac{\dot{f}}{f}\right)A_{0}\right].
\end{equation}
 While the integrability condition gives
\begin{equation}
3H\left(\partial_{i}\partial_{i}A_{0}-\partial_{i}\dot{A}_{i}\right)+\frac{\left(am\right)^{2}}{f}\left(2\frac{\dot{m}}{m}A_{0}+\dot{A}_{0}-a^{-2}\partial_{i}A_{i}\right)=0.
\end{equation}
 Combining these three equations we find
\begin{equation}
\ddot{A}_{i}+\left(H+\frac{\dot{f}}{f}\right)\dot{A}_{i}-a^{-2}\partial_{j}\partial_{j}A_{i}+\frac{m^{2}}{f}A_{i}-a^{-1}\frac{\dot{h}}{f}\epsilon^{ijl}\partial_{j}A_{l}=\left(\frac{\dot{f}}{f}-2\frac{\dot{m}}{m}-2H\right)\partial_{i}A_{0}.\label{eq:integrb}
\end{equation}
 One can immediately notice that axial term changes only the equations
of motion for the spatial components (and not the temporal one) of
the vector field by adding the last term on the LHS. Also, as this
term is proportional to the derivative of the vector field, the axial
term does not have any effect on the homogeneous values of the vector
field components.

As noted in Refs.~\cite{vecurv,sugravec}
the vector field $A_{\mu}$, which enters the Lagrangian, is defined
with respect to the comoving coordinates. While the physical vector
field is $\left(A_{0},A_{i}/a\right)$ for our choice of the metric.
Thus, spatial components of the canonically normalized, physical vector
field are given by
\begin{equation}
W_{i}=\sqrt{f}\frac{A_{i}}{a}.
\end{equation}

To study perturbations $\delta W_{i}$ of $W_{i}$ we go to the momentum
space and write Fourier modes of $\delta W_{i}$ as
\begin{equation}
\delta W_{i}\left(\mathbf{k},t\right)=e_{i}^{\lambda}\left(\hat{\mathbf{k}}\right)w_{\lambda}\left(k,t\right),\label{eq:w-decomp}
\end{equation}
 where summation over $\lambda=`\mathrm{L}',\,`\mathrm{R}'\;\mathrm{and}\;`\|'$
is assumed. Vectors $\boldsymbol{e}^{\mathrm{L}}$, $\boldsymbol{e}^{\mathrm{R}}$
and $\boldsymbol{e}^{\|}$ are three circular polarization vectors
with $\mathbf{k}\cdot\boldsymbol{e}^{\mathrm{L}}=\mathbf{k}\cdot\boldsymbol{e}^{R}=\mathbf{k}\times\boldsymbol{e}^{\|}=0$,
$\mathbf{k}\cdot\boldsymbol{e}^{\|}=k$ and $\mathbf{k}\times\boldsymbol{e}^{\mathrm{L}}=-ik\boldsymbol{e}^{\mathrm{L}}$,
$\mathbf{k}\times\boldsymbol{e}^{\mathrm{R}}=ik\boldsymbol{e}^{\mathrm{R}}$.
Using this and Eq.~\eqref{eq:w-decomp} one easily notices that the
axial term does not affect the longitudinal component of the perturbation.
The equation of motion of this component is exactly the same as studied
in Ref.~\cite{varkin}, 
where it is shown that
a flat perturbation spectrum is achieved if Eq.~\eqref{eq:f-m-scaling}
holds, giving
\begin{equation}
\mathcal{P}_{\|}=\left(\frac{3H}{M}\right)^{2}\left(\frac{H}{2\pi}\right)^{2}.
\end{equation}
 In this equation $M$ is the effective mass of the vector field 
\begin{equation}
M^{2}\equiv\frac{m^{2}}{f}\propto a^{3\pm3},\label{eq:M-def}
\end{equation}
 where $M=\mathrm{constant}$ for $f\propto a^{2}$.

The transverse polarizations, however, are changed by the axial term.
Modified equations of motion for the physical, canonically normalised
vector field transverse modes are given by
\begin{equation}
\ddot{w}_{\mathrm{R},\mathrm{L}}+3H\dot{w}_{\mathrm{R},\mathrm{L}}+\left[\left(\frac{k}{a}\right)^{2}+M^{2}\pm\frac{k}{a}\frac{\dot{h}}{f}\right]w_{\mathrm{R},\mathrm{L}}=0,\label{eq:EoM-wLR}
\end{equation}
 where $w_{\mathrm{R},\mathrm{L}}\left(k,t\right)$ are functions
of a wave-number $k$ and we also used Eq.~\eqref{eq:f-m-scaling}.
The plus sign in the last term of this equation is for the right-handed
polarization. Let us denote 
\begin{equation}
\anm^{2}\equiv\frac{k}{a}\frac{\left|\dot{h}\right|}{f}.\label{eq:N-def}
\end{equation}
 Then Eq.~\eqref{eq:EoM-wLR} can be written as
\begin{equation}
\ddot{w}_{\pm}+3H\dot{w}_{\pm}+\left[\left(\frac{k}{a}\right)^{2}+M^{2}\pm\anm^{2}\right]w_{\pm}=0.\label{eq:EoM-wpm}
\end{equation}
 If $\dot{h}\left(t\right)$ is positive during inflation, subscripts
of $w$ are understood as $`+'\equiv\mathrm{R}$ and $`-'\equiv\mathrm{L}$.
If, instead, $\dot{h}\left(t\right)$ is negative, then $`+'\equiv\mathrm{L}$
and $`-'\equiv\mathrm{R}$.

It is difficult to find a general solution of Eq.~\eqref{eq:EoM-wpm}.
Thus we solve this equation in the regimes where each of the term
proportional to $w_{\pm}$ is dominant and then match these solutions
together.

\subsection{Initial conditions \label{sub:Bunch-Davies}}

Initial conditions for each perturbation mode are set by assuming
that deep within the horizon the $k/a$ term in Eq.~\eqref{eq:EoM-wpm}
dominates. Effectively this means that such a mode describes a free
quantum field and one can set Bunch-Davies initial conditions (see
e.g. Ref.~\cite{book}
)
\begin{equation}
w_{\mathrm{vac}}=\frac{a^{-1}}{\sqrt{2k}}\mathrm{e}^{ik/aH}.
\end{equation}
 It is easy to see that with the above initial conditions modes $w_{\pm}$
and their derivatives are of the form
\begin{eqnarray}
w_{\pm} & = & -a^{-3/2}\frac{1}{2}\sqrt{\frac{\pi}{H}}\mathcal{H}_{3/2}^{\left(1\right)}\left(\frac{k}{aH}\right),\label{eq:w-large-p}\\
\dot{w}_{\pm} & = & a^{-3/2}\frac{1}{2}\sqrt{\frac{\pi}{H}}\frac{k}{a}\mathcal{H}_{1/2}^{\left(1\right)}\left(\frac{k}{aH}\right),\label{eq:wd-large-p}
\end{eqnarray}
 where $\mathcal{H}^{\left(1\right)}$ are Hankel functions of the
first kind.

\subsection{Vector Field Perturbation Spectrum \label{sub:Dom-Q}}

Solutions of Eq.~\eqref{eq:EoM-wpm} with the dominant $M^{2}$ term
can be found in Ref.~\cite{varkin}.
One could also consider the case $\anm\sim M$. But cosmological scales
span seven orders of magnitude in $k$. Because $\anm\propto\sqrt{k}$,
$\anm\sim M$ can apply only to a very narrow range of scales of interest,
and we do not consider this case. In this section, instead, we study
a situation when the dominant term is $\anm^{2}$, i.e. 
\begin{equation}
\ddot{w}_{\pm}+3H\dot{w}_{\pm}\pm\anm^{2}w_{\pm}=0.\label{eq:EoM-wpm-Q}
\end{equation}
 For this we assumed the power law ansatz for $\dot{h}$ allowing
us to write
\begin{equation}
\anm\propto a^{c}.\label{eq:N-ansatz}
\end{equation}
 A general solution of a second order differential equation has two
constants. We determine these constants by requiring that the mode
functions and their derivatives in Eqs.~\eqref{eq:w-large-p} and
\eqref{eq:wd-large-p} match to the solutions of Eq.~\eqref{eq:EoM-wpm-Q}
when the scale factor is $a=\x a$, where
\begin{equation}
\frac{k}{\x a\left(k\right)}=\anm\left(\x a,k\right).
\end{equation}

If $\anm$ is a power law function, Eq.~\eqref{eq:EoM-wpm-Q} is
a Bessel equation. For the $w_{+}$ mode the solution of this equation
is
\begin{eqnarray}
w_{+} & = & a^{-3/2}\left[C_{1}^{+}J_{\nu}\left(\frac{\anm}{\left|c\right|H}\right)+C_{2}^{+}Y_{\nu}\left(\frac{\anm}{\left|c\right|H}\right)\right],\label{eq:wp-N-gen}\\
\dot{w}_{+} & = & -a^{-3/2}\anm\left[C_{1}^{+}J_{\nu+\s c}\left(\frac{\anm}{\left|c\right|H}\right)+C_{2}^{+}Y_{\nu+\s c}\left(\frac{\anm}{\left|c\right|H}\right)\right],\label{eq:wpd-N-gen}
\end{eqnarray}
 where $J_{\nu}$ and $Y_{\nu}$ are Bessel functions of the first
and second kind respectively of order
\begin{equation}
\nu\equiv\frac{3}{2\left|c\right|},
\end{equation}
 and $\s c$ is a signature of $c$, i.e. $\s c=\pm1$ if $\anm$
is an increasing or decreasing function respectively. By matching
these solutions to Eqs.~\eqref{eq:w-large-p} and \eqref{eq:wd-large-p}
we find the constants $\Cpi$ and $\Cpii$
\begin{eqnarray}
C_{1}^{+} & = & \frac{1}{2}\sqrt{\frac{\pi}{H}}\frac{\mathcal{H}_{1/2}Y_{\nu}-\mathcal{H}_{3/2}Y_{\nu+}}{J_{\nu}Y_{\nu+}-J_{\nu+}Y_{\nu}},\label{eq:Cp1-gen}\\
C_{2}^{+} & = & -\frac{1}{2}\sqrt{\frac{\pi}{H}}\frac{\mathcal{H}_{1/2}J_{\nu}-\mathcal{H}_{3/2}J_{\nu+}}{J_{\nu}Y_{\nu+}-J_{\nu+}Y_{\nu}},\label{eq:Cp2-gen}
\end{eqnarray}
 where for brevity we used the following notation: $\mathcal{H}_{\nu}\equiv\mathcal{H}_{\nu}^{\left(1\right)}\left(k/\x aH\right)$,
$J_{\nu}\equiv J_{\nu}\left(k/\left|c\right|\x aH\right)$, $\nu+\equiv\nu+\s c$
and similarly for $Y_{\nu}$.

In the same way we can find the solutions for the $w_{-}$ mode
\begin{eqnarray}
w_{-} & = & a^{-3/2}\left[C_{1}^{-}\mathcal{I}_{\nu}\left(\frac{\anm}{\left|c\right|H}\right)+C_{2}^{-}\mathcal{K}_{\nu}\left(\frac{\anm}{\left|c\right|H}\right)\right],\label{eq:wm-N-dom}\\
\dot{w}_{-} & = & \s ca^{-3/2}\anm\left[C_{1}^{-}\mathcal{I}_{\nu+\s c}\left(\frac{\anm}{\left|c\right|H}\right)-C_{2}^{-}\mathcal{K}_{\nu+\s c}\left(\frac{\anm}{\left|c\right|H}\right)\right],\label{eq:wmd-N-dom}
\end{eqnarray}
 where $\mathcal{I}_{\nu}$ and $\mathcal{K}_{\nu}$ are hyperbolic
Bessel functions. The constants $\Cmi$ and $\Cmii$ are found again
by matching the above solutions with Eqs.~\eqref{eq:w-large-p} and
\eqref{eq:wd-large-p}
\begin{eqnarray}
C_{1}^{-} & = & \frac{1}{2}\sqrt{\frac{\pi}{H}}\frac{\s c\mathcal{H}_{1/2}\mathcal{K}_{\nu}-\mathcal{\mathcal{H}}_{3/2}\mathcal{K}_{\nu+}}{\mathcal{K}_{\nu}\mathcal{I}_{\nu+}+\mathcal{K}_{\nu+}\mathcal{I}_{\nu}},\label{eq:Cm1-gen}\\
C_{2}^{-} & = & -\frac{1}{2}\sqrt{\frac{\pi}{H}}\frac{\s c\mathcal{H}_{1/2}\mathcal{I}_{\nu}+\mathcal{H}_{3/2}\mathcal{I}_{\nu+}}{\mathcal{K}_{\nu}\mathcal{I}_{\nu+}+\mathcal{K}_{\nu+}\mathcal{I}_{\nu}},\label{eq:Cm2-gen}
\end{eqnarray}
 where $\mathcal{I}_{\nu}\equiv\mathcal{I}_{\nu}\left(k/\left|c\right|\x aH\right)$
and the same for $\mathcal{K}_{\nu}$.

\begin{figure}
\begin{centering}
\vspace{-4cm}
\includegraphics[width=100mm]{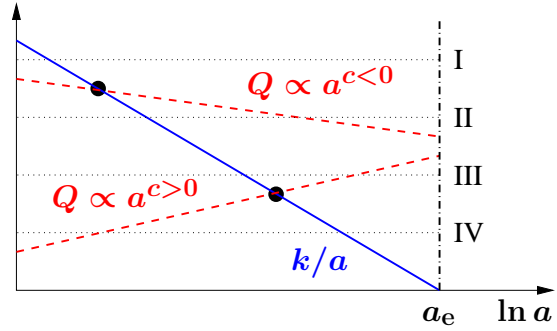}
\vspace{-4cm}
\par\end{centering}

\caption{A log-log graph illustrating a possible evolution of the physical
momentum $k/a$ and the function $\anm\propto a^{c}$ during inflation
(for a fixed $k$). Horizontal lines represent four possible cases
for the value of the Hubble parameter. The end of inflation at 
\mbox{$a=a_{\rm e}$}
is depicted by a vertical dot-dashed line. The mode $k$ exits the horizon
when the blue (solid) line crosses the dotted line. The four possible
cases are the following. In the Case I the function $\anm\ll H$ during
inflation, no matter if it is increasing of decreasing. The Case II
is when a function $\anm\gg H$ before the mode exits the horizon,
but becomes smaller than $H$ towards the end of inflation. This can
only happen if $\anm$ is a decreasing function of time. The Case
III depicts a situation, when $\anm\ll H$ before horizon exit, but
$\e{\anm}\gg H$ at the end of inflation. This only is possible if
$\anm$ is an increasing function of time. Finally, in the Case IV
$\anm\gg H$ for both increasing and decreasing $\anm$ during inflation.
The black dots in the graph highlights the moment when $k/\x aH=\anm$.
\label{fig:p-and-N-evolution}}

\end{figure}

The perturbation power spectrum for each mode is given by
\begin{equation}
\mathcal{P}_{w_{\pm}}=\lim_{\frac{k}{\e aH}\rightarrow0}\frac{k^{3}}{2\pi^{2}}\left|w_{\pm}\right|^{2}.\label{eq:P-def}
\end{equation}
 In the $k/\e aH\rightarrow0$ limit we consider two possible values
of the function $\anm$: either $\e{\anm}/H\ll1$ or $\e{\anm}/H\gg1$,
where subscript `$\mathrm{e}$' denotes values at the end of inflation.
In each case, the function $\anm$ can be decreasing or increasing.
Thus for each value of $\e{\anm}$ we have two possible values of
$k/\x a$ - larger and smaller than $\e{\anm}$ - giving us four possibilities
in total (see Fig.~\ref{fig:p-and-N-evolution}). We study these
four possibilities below and find the perturbation power spectrum
in each case.

%

\subsubsection{The Case I\label{sub:Case-I}}

This corresponds to the case when the inflationary Hubble parameter
is represented by the uppermost dotted line in Fig.~\ref{fig:p-and-N-evolution},
i.e. $k/\x aH\ll1$ and $\e{\anm}/H\ll1$. In this regime the $\anm$
term can be increasing as well as decreasing, i.e. $c>0$ or $c<0$.
We find the constants $\Cpi$ and $\Cpii$ by expanding Eqs.~\eqref{eq:Cp1-gen}
and \eqref{eq:Cp2-gen} for small values of $k/\x aH$
\begin{eqnarray}
C_{1}^{+} & = & -\frac{i\s c}{3}\Gamma\left(\nu+1\right)\sqrt{\frac{k}{2\x aH^{2}}}\left(\frac{k}{2\left|c\right|\x aH}\right)^{-\nu}s_{1}^{+},\label{eq:pC1-small}\\
C_{2}^{+} & = & -\frac{i\s c}{3}\frac{\pi}{\Gamma\left(\nu\right)}\sqrt{\frac{k}{2\x aH^{2}}}\left(\frac{k}{2\left|c\right|\x aH}\right)^{\nu}s_{2}^{+},\label{eq:pC2-small}
\end{eqnarray}
 where
\begin{eqnarray}
s_{1}^{+} & \equiv & 1-\left[3+\left|c\right|\left(\s c-1\right)\right]^{\s c}\left(\frac{k}{\x aH}\right)^{-1-\s c},\\
s_{2}^{+} & \equiv & 1-\left[3+\left|c\right|\left(\s c+1\right)\right]^{-\s c}\left(\frac{k}{\x aH}\right)^{-1+\s c}.
\end{eqnarray}
 Note that, for an increasing $\anm$, $s_{1}^{+}\gg s_{2}^{+}$,
and for the decreasing one $s_{1}^{+}\ll s_{2}^{+}$. In both cases,
they are equal to 
\begin{equation}
s_{1}^{+}\left(c>0\right)=s_{2}^{+}\left(c<0\right)\approx-3\left(\frac{k}{\x aH}\right)^{-2}.\label{eq:sp12}
\end{equation}
 Expanding the Bessel function in Eq.~\eqref{eq:wm-N-dom} with $\e{\anm}/H\ll1$
we find
\begin{equation}
w_{+}=a^{-3/2}\left[C_{1}^{+}\frac{1}{\Gamma\left(\nu+1\right)}\left(\frac{\anm}{2\left|c\right|H}\right)^{\nu}-C_{2}^{+}\frac{\Gamma\left(\nu\right)}{\pi}\left(\frac{\anm}{2\left|c\right|H}\right)^{-\nu}\right].\label{eq:wp-small-N}
\end{equation}
 Inserting Eqs.~\eqref{eq:pC1-small}, \eqref{eq:pC2-small} and taking into account Eq.~\eqref{eq:sp12}, we
finally obtain
\begin{equation}
w_{+}=\frac{i}{\sqrt{2}}Hk^{-3/2}.
\end{equation}
 Using Eq.~\eqref{eq:P-def} we see that the spectrum of $w_{+}$
is flat
\begin{equation}
\mathcal{P}_{w_{+}}^{\left(\mathrm{I}\right)}=\left(\frac{H}{2\pi}\right)^{2}.\label{eq:CI-Pwp}
\end{equation}

We apply the same method for calculating the power spectrum of the
$w_{-}$ mode. In the limit of small $k/\x aH$ constants $\Cmi$
and $\Cmii$ in Eqs.~\eqref{eq:Cm1-gen} and \eqref{eq:Cm2-gen}
become
\begin{eqnarray}
C_{1}^{-} & = & -\frac{i}{3}\Gamma\left(\nu+1\right)\sqrt{\frac{k}{2\x aH^{2}}}\left(\frac{k}{2\left|c\right|\x aH}\right)^{-\nu}s_{1}^{-},\label{eq:mC1-px-small}\\
C_{2}^{-} & = & \frac{i}{3}\frac{2}{\Gamma\left(\nu\right)}\sqrt{\frac{k}{2\x aH^{2}}}\left(\frac{k}{2\left|c\right|\x aH}\right)^{\nu}s_{2}^{-},\label{eq:mC2-px-small}
\end{eqnarray}
 where $s_{1}^{-}$ and $s_{2}^{-}$ are given by 
\begin{eqnarray}
s_{1}^{-} & \equiv & \s c-\left[3+\left|c\right|\left(\s c-1\right)\right]^{\s c}\left(\frac{k}{\x aH}\right)^{-1-\s c},\\
s_{2}^{-} & \equiv & \s c+\left[3+\left|c\right|\left(\s c+1\right)\right]^{-\s c}\left(\frac{k}{\x aH}\right)^{-1+\s c}.
\end{eqnarray}
 One can see again that for $c>0$, $s_{1}^{-}\gg s_{2}^{-}$ and
\emph{vice versa}. Also for each of these cases
\begin{equation}
s_{1}^{-}\left(c>0\right)=s_{2}^{-}\left(c<0\right)\approx-\s c3\left(\frac{k}{\x aH}\right)^{-2}.\label{eq:sm12}
\end{equation}

Next, expanding Eq.~\eqref{eq:wm-N-dom} for small $\e{\anm}/\left|c\right|H$
we find
\begin{equation}
w_{-}=a^{-3/2}\left[\frac{\Cmi}{\Gamma\left(\nu+1\right)}\left(\frac{\e{\anm}}{2\left|c\right|H}\right)^{\nu}+\Cmii\frac{\Gamma\left(\nu\right)}{2}\left(\frac{\e{\anm}}{2\left|c\right|H}\right)^{-\nu}\right].\label{eq:wm-small-N}
\end{equation}
 It is easy to find that, using Eq.~\eqref{eq:sm12} and expressions
for $\Cmi$ and $\Cmii$ in Eqs.~\eqref{eq:mC1-px-small} and \eqref{eq:mC2-px-small},
the mode function $w_{-}$ becomes
\begin{equation}
w_{-}=\frac{i\s c}{\sqrt{2}}Hk^{-3/2}.
\end{equation}
 Thus the power spectrum is also flat and equal to
\begin{equation}
\mathcal{P}_{w_{-}}^{\left(\mathrm{I}\right)}=\left(\frac{H}{2\pi}\right)^{2}.\label{eq:CI-Pwm}
\end{equation}

As we can see from Eqs.~\eqref{eq:CI-Pwp} and \eqref{eq:CI-Pwm},
the power spectra for both transverse modes are identical at first
order in $\e{\anm}/H$ and are equal to the one of the light scalar
field. This is not surprising as the equations of motion for $w_{+}$
and $w_{-}$ in Eq.~\eqref{eq:EoM-wpm-Q} reduces to the one of the
light scalar field. The spectral tilt of $\mathcal{P}_{w_{\pm}}$,
however, will differ from the scalar field case as the $\anm\left(t,k\right)$
is a function of time and momentum.

To find the spectral tilt of $\mathcal{P}_{w_{\pm}}$ we expand the
Bessel functions to the second order in $\e{\anm}/H$. Let us consider
$\mathcal{P}_{w_{+}}$ first. Then
\begin{eqnarray}
J_{\nu}\left(\frac{\anm}{\left|c\right|H}\right) & = & \frac{1}{\Gamma\left(1+\nu\right)}\left(\frac{\anm}{2\left|c\right|H}\right)^{\nu}\left[1-\frac{1}{1+\nu}\left(\frac{\anm}{2\left|c\right|H}\right)^{2}\right]+\mathcal{O}\left[\left(\frac{\anm}{H}\right)^{\nu+4}\right],\\
Y_{\nu}\left(\frac{\anm}{\left|c\right|H}\right) & = & -\frac{\Gamma\left(\nu\right)}{\pi}\left(\frac{\anm}{2\left|c\right|H}\right)^{-\nu}\left[1-\frac{1}{1-\nu}\left(\frac{\anm}{2\left|c\right|H}\right)^{2}\right]+\mathcal{O}\left[\left(\frac{\anm}{H}\right)^{\nu+4}\right].
\end{eqnarray}
 The above expression for $Y_{\nu}$ is valid only for $\nu>1$. However,
the $Y_{\nu}$ term in Eq.~\eqref{eq:wp-N-gen} is dominant only
when this condition is satisfied. From the above, it is easy to infer
that, to second order in $\e Q/H$, the power spectrum of $w_{+}$
is 
\begin{equation}
\mathcal{P}_{w_{+}}=\left(\frac{H}{2\pi}\right)^{2}\left[1-\frac{4c}{3+2c}\left(\frac{\e{\anm}}{2\left|c\right|H}\right)^{2}\right].\label{eq:CI-Pwp-2nd}
\end{equation}

Performing the same expansion for the hyperbolic Bessel functions
we find
\begin{eqnarray}
\mathcal{I}_{\nu}\left(\frac{\anm}{\left|c\right|H}\right) & = & \frac{1}{\Gamma\left(1+\nu\right)}\left(\frac{\anm}{2\left|c\right|H}\right)^{\nu}\left[1+\frac{1}{1+\nu}\left(\frac{\anm}{2\left|c\right|H}\right)^{2}\right]+\mathcal{O}\left[\left(\frac{\anm}{H}\right)^{\nu+4}\right],\\
\mathcal{K}_{\nu}\left(\frac{\anm}{\left|c\right|H}\right) & = & \frac{\Gamma\left(\nu\right)}{2}\left(\frac{\anm}{2\left|c\right|H}\right)^{-\nu}\left[1+\frac{1}{1-\nu}\left(\frac{\anm}{2\left|c\right|H}\right)^{2}\right]+\mathcal{O}\left[\left(\frac{\anm}{H}\right)^{\nu+4}\right].
\end{eqnarray}
 With this result, the power spectrum of $w_{-}$ becomes
\begin{equation}
\mathcal{P}_{w_{-}}=\left(\frac{H}{2\pi}\right)^{2}\left[1+\frac{4c}{3+2c}\left(\frac{\e{\anm}}{2\left|c\right|H}\right)^{2}\right].\label{eq:CI-Pwm-2nd}
\end{equation}

The two expressions for $\mathcal{P}_{w_{\pm}}$ above can be combined
into one result
\begin{equation}
\mathcal{P}_{w_{\pm}}=\left(\frac{H}{2\pi}\right)^{2}\left[1\mp\varepsilon\frac{k}{H}\right],
\end{equation}
 where
\begin{equation}
\varepsilon\equiv\frac{\e a^{2c}}{c\left(3+2c\right)}\,\frac{\dot{h}_{0}}{Hf_{0}}.
\end{equation}
 As it is clear from the above, the spectral dependence of the power
spectra of $w_{+}$ and $w_{-}$ are the same but with opposite signs.
The power spectrum of $\zeta$ in Eq.~\eqref{eq:Pz-def}, on the
other hand, is proportional only to the arithmetic average of $\mathcal{P}_{w_{+}}$
and $\mathcal{P}_{w_{-}}$ in which momentum dependence cancels out.
Thus the vector field in this case may influence the spectral tilt
of $\mathcal{P}_{\zeta}$ only at even higher order in $\e{\anm}/H\ll1$,
which probably makes it undetectably small.

Spectral tilts of $\mathcal{P}_{w_{\pm}}$ do not modify $\fnl$ significantly
too. As can be seen from the definitions of the parameters $p$ and
$q$ in Eq.~\eqref{eq:pq-def} and the expressions for $\fnl$ in
Eqs.~\eqref{eq:fNL-equil} - \eqref{eq:fNL-flat}, the above spectral
dependence influences $\fnl$ only through the parameter $q$. In
this case, however, $q\left(k\right)\ll1$.

\subsubsection{The Case II\label{sub:Case-II}}

Case II corresponds to when the $\anm$ term dominates the $k/a$
one in Eq.~\eqref{eq:EoM-wpm} before horizon exit, but at the end
of inflation $\e{\anm}\ll H$. This is possible only if $\anm$ is
a decreasing function of time, that is $c<0$. As discussed in subsection
\ref{sub:Bunch-Davies}, however, to set initial Bunch-Davies conditions
the $k/a$ term in Eq.~\eqref{eq:EoM-wpm} must dominate initially.
Therefore the $Q$ term must decrease slower than $k/a$, putting
the lower bound on $c>-1$.

In the limit $k/\x aH\gg1$, the constants $C_{1}^{+}$ and $C_{2}^{+}$
in Eqs.~\eqref{eq:Cp1-gen} and \eqref{eq:Cp2-gen} become
\begin{equation}
C_{2}^{+}=-i\s cC_{1}^{+}=-\frac{i\s c}{2}\sqrt{\frac{\pi}{\left|c\right|H}}\mathrm{e}^{i\psi},\label{eq:Cp1-Cp2-large-px}
\end{equation}
 where 
\begin{equation}
\psi\equiv\frac{k}{\x aH}\left(1+\frac{1}{c}\right)-\s c\frac{2\nu+1}{4}\pi.
\end{equation}

The equation for $w_{+}$ in the limit $\e{\anm}/H\ll1$ is given
in Eq.~\eqref{eq:wp-small-N}, with the constants $C_{1}^{+}$ and
$\Cpii$ determined by Eq.~\eqref{eq:Cp1-Cp2-large-px}. As $\nu$
is positive by definition, the second term on the RHS of Eq.~\eqref{eq:wp-small-N}
dominates. Inserting it into Eq.~\eqref{eq:P-def}, we find
\begin{equation}
\mathcal{P}_{w_{+}}^{\left(\mathrm{II}\right)}=\left(2\left|c\right|\right)^{2\nu-1}\frac{\Gamma^{2}\left(\nu\right)}{\pi}\left(\frac{\dot{h}}{Hf}\right)^{-\nu}\left(\frac{k}{aH}\right)^{3-\nu}\left(\frac{H}{2\pi}\right)^{2}.
\end{equation}
 One notices that with $c=-1/2$, $\nu=3$ and the power spectrum
of $w_{+}$ is flat
\begin{equation}
\mathcal{P}_{w_{+}}^{\left(\mathrm{II}\right)}=\frac{4}{\pi}\left(\frac{\dot{h}}{Hf}\right)^{-3}\left(\frac{H}{2\pi}\right)^{2}.\label{eq:CII-P-wp}
\end{equation}

Let's consider the $w_{-}$ mode next. In the limit $k/\x aH\gg1$
the constants $\Cmi$ and $\Cmii$ become
\begin{eqnarray}
\left|\Cmi\right| & = & \sqrt{\frac{\pi}{2\left|c\right|H}}\mathrm{e}^{-k/\x a\left|c\right|H},\label{eq:Cm1-larpe-px}\\
\left|\Cmii\right| & = & \frac{1}{\sqrt{2\pi\left|c\right|H}}\mathrm{e}^{k/\x a\left|c\right|H},\label{eq:Cm2-large-px}
\end{eqnarray}
 and in the limit $\e{\anm}/H\ll1$ the mode function $w_{-}$ is
given in Eq.~\eqref{eq:wm-small-N}. Because $\e{\anm}/H\ll1$ and
$\left|\Cmii\right|\gg\left|\Cmi\right|$ the second term dominates
and its contribution to the power spectrum is
\begin{equation}
\mathcal{P}_{w_{-}}^{\left(\mathrm{II}\right)}=\mathcal{P}_{w_{+}}^{\left(\mathrm{II}\right)}\frac{1}{2}\mathrm{e}^{\Theta^{\left(\mathrm{II}\right)}},\label{eq:CII-P-wm}
\end{equation}
 where
\begin{equation}
\Theta^{\left(\mathrm{II}\right)}\equiv\frac{2}{\left|c\right|}\left(\frac{\dot{h}}{Hf}\right)^{\frac{1}{2\left(c+1\right)}}\left(\frac{k}{aH}\right)^{\frac{1+2c}{2\left(c+1\right)}}.\label{eq:Th-II-def}
\end{equation}
 Here again one can easily notice that, with $c=-1/2$, the exponential
amplification of the power spectrum $\mathcal{P}_{w_{-}}^{\left(\mathrm{II}\right)}$
is scale independent 
\begin{equation}
\Theta^{\left(\mathrm{II}\right)}=4\frac{\dot{h}}{Hf}.
\end{equation}

\subsubsection{The Case III\label{sub:Case-III}}

Case III in Figure~\ref{fig:p-and-N-evolution} corresponds to the
situation when the $\anm$ term takes over the $k/a$ term in Eq.~\eqref{eq:EoM-wpm}
after horizon exit, but $\anm$ is a growing function of time and
becomes larger than $H$ at the end of inflation. We find the power
spectra of $w_{+}$ and $w_{-}$ by the same method as in the previous
two subsections.

Since Case III corresponds to $k/\x aH\ll1$, constants $\Cpi$ and
$\Cpii$ in this limit are given in Eqs.~\eqref{eq:pC1-small} and
\eqref{eq:pC2-small}. While in the limit $\e{\anm}/H\gg1$, Eq.~\eqref{eq:wp-N-gen}
for $w_{+}$ becomes 
\begin{equation}
w_{+}=a^{-3/2}\sqrt{\frac{2\left|c\right|H}{\pi\e{\anm}}}\left[\Cpi\cos\left(\frac{\e{\anm}}{\left|c\right|H}-\frac{2\nu+1}{4}\pi\right)+\Cpii\sin\left(\frac{\e{\anm}}{\left|c\right|H}-\frac{2\nu+1}{4}\pi\right)\right].\label{eq:wp-largeN}
\end{equation}
 For an increasing $\anm$, i.e. $c>0$, $\left|\Cpi\right|\gg\left|\Cpii\right|$
and the first term in the above equation dominates. Using this and
Eqs.~\eqref{eq:pC1-small} we find
\begin{equation}
\mathcal{P}_{w_{+}}^{\left(\mathrm{III}\right)}=\left(2\left|c\right|\right)^{1+2\nu}\frac{\Gamma^{2}\left(\nu+1\right)}{\pi}\left(\frac{k}{aH}\frac{\e{\dot{h}}}{H\e f}\right)^{-\frac{c+3}{c}}\left(\frac{H}{2\pi}\right)^{2}\cos^{2}\Theta^{\left(\mathrm{III}\right)},\label{eq:Pwp-III}
\end{equation}
 where
\begin{equation}
\Theta^{\left(\mathrm{III}\right)}\equiv\frac{\e{\anm}}{\left|c\right|H}-\frac{2\nu+1}{4}\pi.\label{eq:Th-III-def}
\end{equation}
 As $c>0$ the power spectrum $\mathcal{P}_{w_{+}}$ in this case
cannot be scale invariant.

The integration constants of the $w_{-}$ mode for $k/\x aH\ll1$
are given in Eqs.~\eqref{eq:mC1-px-small} and \eqref{eq:mC2-px-small}.
Expanding Eq.~\eqref{eq:wm-N-dom} in the limit $\e{\anm}/H\gg1$
we find
\begin{equation}
w_{-}=a^{-3/2}\sqrt{\frac{\left|c\right|H}{2\pi\e{\anm}}}\left(\Cmi\mathrm{e}^{\e{\anm}/\left|c\right|H}+\pi\Cmii\mathrm{e}^{-\e{\anm}/\left|c\right|H}\right).\label{eq:wm-largeN}
\end{equation}
 For increasing $\anm$, the constants are $\left|\Cmi\right|\gg\left|\Cmii\right|$
and the first term dominates. Using this we find
\begin{equation}
\mathcal{P}_{w_{-}}^{\left(\mathrm{III}\right)}=\mathcal{P}_{w_{+}}^{\left(\mathrm{III}\right)}\frac{1}{4}\mathrm{e}^{2\e{\anm}/\left|c\right|H}.
\end{equation}
 Because $\e{\anm}/H\gg1$ is a function of a wave-number $k$, we
see that the power spectrum of $w_{-}$ features an exponential, scale
dependent amplification factor.

\subsubsection{The Case IV\label{sub:Case-IV}}

In the Case IV, $\anm$ is always larger than the Hubble parameter
during inflation. Such scenario corresponds to $k/\x a\gg H$ and
$\e{\anm}\gg H$. We have already calculated the integration constants
and the equations for the mode functions in the previous subsections.
The expression for $w_{+}$ mode in the large $\e{\anm}$ limit is
given in Eq.~\eqref{eq:wp-largeN}. While the constants $\Cpi$ and
$\Cpii$ in the large $k/H\x a$ limit can be found in Eq.~\eqref{eq:Cp1-Cp2-large-px}.
With these equations it is easy to find the following solution for
$w_{+}$
\begin{equation}
w_{+}=a^{-3/2}\Cpi\sqrt{\frac{2\left|c\right|H}{\pi\e{\anm}}}\mathrm{e}^{i\s c\Theta^{\left(\mathrm{III}\right)}},
\end{equation}
 where the phase factor $\Theta^{\left(\mathrm{III}\right)}$ is defined
in Eq.~\eqref{eq:Th-III-def}. Substituting this into Eq.~\eqref{eq:P-def},
we find
\begin{equation}
\mathcal{P}_{w_{+}}^{\left(\mathrm{IV}\right)}=\left(\frac{\dot{h}}{Hf}\right)^{-1/2}\left(\frac{k}{aH}\right)^{5/2}\left(\frac{H}{2\pi}\right)^{2}.\label{eq:CIV-P-wp}
\end{equation}
 As one can see this spectrum is very blue.

In an analogous way, we can compute the power spectrum of the $w_{-}$
mode, which in the large $\e{\anm}$ limit is given in Eq.~\eqref{eq:wm-largeN}.
The integration constants $\Cmi$ and $\Cmii$ were calculated in
Eqs.~\eqref{eq:Cm1-larpe-px} and \eqref{eq:Cm2-large-px}. Putting
these expressions together we find 
\begin{equation}
w_{-}=\frac{a^{-3/2}}{2\sqrt{\e{\anm}}}\left[\mathrm{e}^{\left(\e{\anm}-k/\x a\right)/\left|c\right|H}+\mathrm{e}^{-\left(\e{\anm}-k/\x a\right)/\left|c\right|H}\right].
\end{equation}
 When $\anm$ is an increasing function of time, i.e. $c>0$, then
$\e{\anm}\gg k/\x a$ and the first term in the above expression dominates.
While, for $c<0$, the second term dominates. With this solution,
the spectrum of $w_{-}$ becomes
\begin{equation}
\mathcal{P}_{w_{-}}^{\left(\mathrm{IV}\right)}=\mathcal{P}_{w_{+}}^{\left(\mathrm{IV}\right)}\frac{1}{2}\mathrm{e}^{2\left(\e{\anm}-k/\x a\right)/cH}.\label{eq:CIV-P-wm}
\end{equation}
 Thus, again in this case we find that the power spectrum of $w_{-}$
has an exponential $k$-dependent amplification.

\subsubsection{Logarithmic $h$}

The last two cases - logarithmic $h$ and constant $Q$ - are not
shown in Figure~\ref{fig:p-and-N-evolution}. In the first, logarithmic
$h$ case the time derivative of $h$ is constant during (quasi) de
Sitter inflation:
\begin{equation}
\dot{h}=Hh_{0}\approx\mathrm{constant},
\end{equation}
 where $h_{0}$ is the initial value of $h$. From the definition
of the function $\anm$ in Eq.~\eqref{eq:N-def} we notice that
\begin{equation}
\anm\propto a^{-\left(1+\alpha\right)/2},\label{eq:N-log-h}
\end{equation}
 where $\alpha\equiv-1\pm3$ is the scaling of the kinetic function
$f\propto a^{\alpha}$. While from the definition of $c$ in Eq.~\eqref{eq:N-ansatz}
we can see that the logarithmic scaling of $h$ corresponds to
\begin{equation}
c=\pm3/2,
\end{equation}
 where $c=-3/2$ is for $\alpha=2$. With negative $c$ the function
$\anm$ is decreasing. But in order to be able to impose the Bunch-Davies
initial conditions $\anm$ has to be decreasing slower than $k/a$.
Thus $c=-3/2$ is unviable. However, for $\alpha=-4$, $c=3/2$ and
we can use the results of subsections~\ref{sub:Case-I}, \ref{sub:Case-III}
and \ref{sub:Case-IV} to deduce the power spectrum.

\subsubsection{Constant $\anm$\label{sub:Const-N}}

For the constant $\anm$ case, the equations of motion of both modes
$w_{+}$ and $w_{-}$ in Eq.~\eqref{eq:EoM-wpm} reduce to the familiar
equation of a massive scalar field. The difference, however, is that
the effective {}``mass'' in this case is scale dependent, $\anm\propto k^{1/2}$
and for the $w_{-}$ mode this {}``mass'' is tachyonic. For $\anm\ll H$,
the power spectrum is well known for this type of equation
\begin{equation}
\mathcal{P}_{w_{\pm}}^{\left(\mathrm{const}\right)}=\left(\frac{H}{2\pi}\right)^{2}\left(\frac{k}{2aH}\right)^{3-2\nu_{\pm}},
\end{equation}
 where
\begin{equation}
\nu_{\pm}\equiv\sqrt{\frac{9}{4}\mp\left(\frac{\anm}{H}\right)^{2}}.\label{eq:ni-pm-def}
\end{equation}

We see that in the limit $\anm\ll H$, the spectrum is almost flat.
It is also clear that both modes will have the same spectral tilt,
but of the opposite sign: the spectrum of $w_{+}$ is slightly blue-tilted
and red-tilted for $w_{-}$. However, as discussed in Subsection~\ref{sub:Case-I}
this does not affect the spectral tilt of $\mathcal{P}_{\zeta}$ and
makes a negligible contribution to $\fnl$.

\subsubsection{Summary of the Subsection~\label{sub:Sum-Sec} }

We summarize the results of subsection~\ref{sub:Dom-Q} in Table~\ref{tab:summary}.
There are two parameter spaces for producing a flat perturbation spectrum
for both of transverse modes. First, this can be realised for any
value of $c$ if the $\anm$ term in Eq.~\eqref{eq:EoM-wpm} dominates
the $k/a$ term after modes exit the horizon and if $\anm$ stays
smaller than $H$ until the end of inflation%
\footnote{We assumed here $M\ll\anm$.}. 
Such a vector field can generate statistical anisotropy in the curvature 
perturbation $\zeta$ which is consistent with the observational bounds.
This can be realized, for example, using
the vector curvaton scenario \cite{vecurv}. In this case
the results of the vector curvaton scenario discussed in Ref.~\cite{varkin}
are directly applicable%
\footnote{The only difference are the additional constraints $k/\x aH\ll1$
and $\e{\anm}/H\ll1$.%
}. But as both transverse modes acquire the same perturbation amplitude,
the axial term in the Lagrangian will not have a detectable signature
in the curvature perturbation in this parameter space.

However, in this paper we are also interested in a possibility of producing
$\zeta$ with parity violating statistics. As shown in Sec.~\ref{sub:Case-II}
this can be realized if the function $\anm$ scales as $c=-1/2$ and
if it dominates $k/a$ term in Eq.~\eqref{eq:EoM-wpm} before horizon
exit. Note, however, that $\anm\propto k^{1/2}$. Therefore the above
discussion is valid only for some limited range of $k$ values. Particularly,
$\mathcal{P}_{w_{+}}=\mathcal{P}_{w_{-}}=\left(H/2\pi\right)^{2}$
if $k/\x aH\ll1$ and $\e{\anm}\left(k\right)/H\ll1$. And parity
violating perturbations are realized if the former bound is reversed
and $c=-1/2$. We study the latter case in more detail in Sec.~\ref{sub:Scale-Invariant}.

\begin{table}
\begin{tabular}{|c|c|c|}
\hline 
 & $\mathcal{P}_{w_{+}}$  & $\mathcal{P}_{w_{-}}$\tabularnewline [1ex]
\hline 

Case I \rule{0pt}{3.4ex} & $\left(\frac{H}{2\pi}\right)^{2}$  & $\mathcal{P}_{w_{+}}$\tabularnewline [1ex]
\hline 
Case II \rule{0pt}{4ex}  & $\left(2\left|c\right|\right)^{2\nu-1}\frac{\Gamma^{2}\left(\nu\right)}{\pi}\left(\frac{\e{\dot{h}}}{H\e f}\right)^{-\nu}\left(\frac{k}{\e aH}\right)^{3-\nu}\left(\frac{H}{2\pi}\right)^{2}$  & $\mathcal{P}_{w_{+}}\frac{1}{2}\mathrm{e}^{\Theta^{\left(\mathrm{II}\right)}}$\tabularnewline [2ex]
\hline 
Case III  \rule{0pt}{4.5ex} & $\left(2\left|c\right|\right)^{1+2\nu}\frac{\Gamma^{2}\left(\nu+1\right)}{\pi}\left(\frac{k}{\e aH}\frac{\e{\dot{h}}}{H\e f}\right)^{-\frac{c+3}{c}}\left(\frac{H}{2\pi}\right)^{2}\cos^{2}\Theta^{\left(\mathrm{III}\right)}$  & $\mathcal{P}_{w_{+}}\frac{1}{4}\mathrm{e}^{2\e{\anm}/\left|c\right|H}$\tabularnewline [2ex]
\hline 
Case IV \rule{0pt}{4.5ex} & $\left(\frac{\e{\dot{h}}}{H\e f}\right)^{-1/2}\left(\frac{k}{\e aH}\right)^{5/2}\left(\frac{H}{2\pi}\right)^{2}$  & $\mathcal{P}_{w_{+}}\frac{1}{2}\mathrm{e}^{2\left(\e{\anm}-k/\x a\right)/cH}$\tabularnewline [2ex]
\hline 
$h\propto\ln a$ \rule{0pt}{4ex}  & \multicolumn{2}{c|}{The same as cases I, III and IV with $c=3/2$}\tabularnewline [2ex]
\hline 
$\anm=\mathrm{constant}$ \rule{0pt}{4.5ex} & \multicolumn{2}{c|}{$\mathcal{P}_{w_{\pm}}=\frac{4}{\pi}\Gamma^{2}\left(\nu_{\pm}\right)\left(\frac{H}{2\pi}\right)^{2}\left(\frac{k}{2\e aH}\right)^{3-2\nu_{\pm}}$}\tabularnewline [2ex]
\hline
\end{tabular}

\caption{Summary of subsection~\ref{sub:Dom-Q}. In this table $\Theta^{\left(\mathrm{II}\right)}$
is defined in Eq.~\eqref{eq:Th-II-def}, $\Theta^{\left(\mathrm{III}\right)}$
is defined Eq.~\eqref{eq:Th-III-def} and $\nu_{\pm}$ in Eq.~\eqref{eq:ni-pm-def}.
\label{tab:summary}}

\end{table}

\section{Statistically Anisotropic, Parity Violating Curvature Perturbation
\label{sub:Scale-Invariant}}

\subsection{The Spectrum}

In Subsection~\ref{sub:Case-I} we found that the vector field acquires
a scale invariant perturbation spectrum if the $\anm$ term in Eq.~\eqref{eq:EoM-wpm}
dominates the momentum term $k/a$ after the mode $k$ exits the horizon
and $\e{\anm}\ll H$ at the end of inflation. In this case both modes,
$w_{+}$ and $w_{-}$, acquire perturbation spectrum equal to $\left(H/2\pi\right)^{2}$.
While in such a set-up the scale invariant curvature perturbation
can also be generated, it does not give any signature of parity violation.

A more interesting case is studied in Subsection~\ref{sub:Case-II}.
In this case, we find that the perturbation spectrum of the vector
field is also scale invariant if the $\anm$ term dominates the momentum
one in Eq.~\eqref{eq:EoM-wpm} before the mode exits the horizon
and if $\anm$ is decaying during inflation as $\anm\propto a^{-1/2}$
with $\e{\anm}\ll H$ at the end of inflation. From the definition
of $\anm$ in Eq.~\eqref{eq:N-def} we see that this scaling of $\anm$
corresponds to 
\begin{equation}
\frac{\dot{h}}{f}=\mathrm{constant}.
\end{equation}
 With the power law evolution of $f\propto a^{\alpha}$ the above
condition implies that $h$ also evolves according to the same power
law $h\propto a^{\alpha}$ %
\footnote{Remember that we assume (quasi) de Sitter inflation, i.e. $H\approx\mathrm{constant}$.%
}, where $\alpha=-1\pm3$. In view of this, we can rewrite the Lagrangian
in Eq.~\eqref{eq:Lagrangian} as
\begin{equation}
\mathcal{L}=-\frac{1}{4}f\left(F_{\mu\nu}F^{\mu\nu}+\vartheta F_{\mu\nu}\tilde{F}^{\mu\nu}\right)+\frac{1}{2}m^{2}A_{\mu}A^{\mu},\label{eq:Flat-L}
\end{equation}
 where 
\begin{equation}
\vartheta\equiv\frac{h}{f}=\mathrm{constant}.\label{eq:Flat-theta-def}
\end{equation}

Then a vector field with the Lagrangian in Eq.~\eqref{eq:Flat-L}
will acquire a flat perturbation spectrum if the $\anm^{2}=\left|\alpha\vartheta\right|Hk/a$
term dominates the $\left(k/a\right)^{2}$ term in Eq.~\eqref{eq:EoM-wpm}
before the mode exits the horizon. In the case when $\anm\propto a^{-1/2}$
this happens at a scale factor $\x a$ given by%
\footnote{The moment $k/\x a$ is represented by black dots in Figure~\ref{fig:p-and-N-evolution}. %
} 
\begin{equation}
\frac{k}{H\x a}=\frac{\dot{h}}{Hf}=\left|\alpha\vartheta\right|.\label{eq:Flat-kax}
\end{equation}
 From this we see that the value of $k/\x a$ does not depend neither
on time nor on wave-number $k$.

To calculate the perturbation spectrum of the vector field we used
Bunch-Davies initial conditions in Subsection~\ref{sub:Bunch-Davies}.
These initial conditions are valid in the limit where the curvature
of space-time can be neglected and quantum fields are effectively
described by the quantum theory of free fields. The first requirement
is fulfilled for modes deep within the horizon, while the standard
quantum field theory can be applied for modes which are not too close
to the Planck scale, and the mode can be considered to be of an effectively
free field if the $k/a$ term in Eq.~\eqref{eq:EoM-wpm} dominates.
Using Eq.~\eqref{eq:Flat-kax} we find that these requirements constrain
the value of $\vartheta$ as
\begin{equation}
1\ll\left|\alpha\vartheta\right|\ll\frac{\mpl}{H}.\label{eq:Flat-BD-bound}
\end{equation}

Because the parameter $\alpha$ is of order one, we see that the scale
invariant vector field perturbation spectrum is achieved if the parity
violating constant $\vartheta$ is much larger than unity. Then the
spectra of modes $w_{-}$ and $w_{+}$ from Eqs.~\eqref{eq:CII-P-wp}
and \eqref{eq:CII-P-wm} can be written as

\begin{eqnarray}
\mathcal{P}_{w_{+}} & = & \frac{4}{\pi}\left|\alpha\vartheta\right|^{-3}\left(\frac{H}{2\pi}\right)^{2};\label{eq:Flat-P-wp}\\
\mathcal{P}_{w_{-}} & = & \frac{4}{\pi}\left|\alpha\vartheta\right|^{-3}\left(\frac{H}{2\pi}\right)^{2}\frac{\mathrm{e}^{4\left|\alpha\vartheta\right|}}{2}.\label{eq:Flat-P-wm}
\end{eqnarray}
 Note, that due to Eq.~\eqref{eq:Flat-BD-bound} the spectrum of
$w_{+}$ is suppressed with respect to the standard light scalar field
result.

For the expressions in Eqs.~\eqref{eq:Flat-P-wp} and \eqref{eq:Flat-P-wm}
to be valid, the $\anm$ term must also be much smaller than $H$
by the end of inflation, $\e{\anm}\ll H$. It follows from Eq.~\eqref{eq:Flat-kax}
that this implies
\begin{equation}
\left|\alpha\vartheta\right|\ll\frac{\e aH}{k},\label{eq:Flat-N-ll-H}
\end{equation}
 which must hold at least for cosmological scales. Assuming the observable
inflation to last at least 50 e-folds, this bound is much weaker than
the one in Eq.~\eqref{eq:Flat-BD-bound} for any realistic model
of inflation.

If for the largest $k$ modes this bound is violated, we are back
to the Case IV in Figure~\ref{fig:p-and-N-evolution}, which was
discussed in Subsection~\ref{sub:Case-IV}. As one can see in Eqs.~\eqref{eq:CIV-P-wp}
and \eqref{eq:CIV-P-wm}, the spectra of both modes are very blue
$\mathcal{P}_{w_{\pm}}\propto k^{5/2}$. As the $\mathcal{P}_{w_{-}}$
is also exponentially enhanced by $\exp\left(4\left|\alpha\vartheta\right|\right)$
one may worry about the overproduction of the primordial black holes
in such a scenario. The summary of various cosmological constraints
on the abundance of the primordial black holes can be found in 
Ref.~\cite{carretal}.
On practically all scales the bound correspond to $\mathcal{P}_{\zeta}\lesssim10^{-2}$
for the Gaussian perturbation. If the perturbation is non-Gaussian
then the bound becomes 
\begin{equation}
\mathcal{P}_{\zeta}\left(k_{\mathrm{peak}}\right)\lesssim10^{-3}\;\mathrm{or}\;1,\label{eq:Flat-Pz-BH}
\end{equation}
 where the lower bound is for the positive non-Gaussianity and the
upper bound is for the negative non-Gaussianity 
\cite{lythpbh}.
The contribution of the vector field to the spectrum of $\zeta$ is
$N_{W}^{2}\mathcal{P}_{w_{-}}$, where $N_{W}$ is defined in Eq.~\eqref{eq:dN-formula}.
From COBE normalization this contribution has to be $\lesssim10^{-9}$
for cosmological scales. From Eqs.~\eqref{eq:Flat-P-wp} and \eqref{eq:Flat-P-wm}
$\mathcal{P}_{w_{-}}\gg\mathcal{P}_{w_{+}}$ and thus the spectrum
of $\zeta$ is
\begin{equation}
N_{W}^{2}\frac{2}{\pi}\left|\alpha\vartheta\right|^{-3}\left(\frac{H}{2\pi}\right)^{2}\mathrm{e}^{4\left|\alpha\vartheta\right|}\lesssim10^{-9},\label{eq:Flat-Pz-contr}
\end{equation}
 for cosmological scales.

The spectrum of modes which violate the bound in Eq.~\eqref{eq:Flat-N-ll-H}
is given in Eq.~\eqref{eq:CIV-P-wm}. The blue spectrum peaks at
the largest $k$ value, which corresponds to the horizon size at the
end of inflation. Thus setting $k_{\mathrm{peak}}/\e aH\approx1$
and using Eq.~\eqref{eq:CIV-P-wm} we can write the bound in Eq.~\eqref{eq:Flat-Pz-BH}
as
\begin{equation}
\frac{1}{2}N_{W}^{2}\left|\alpha\vartheta\right|^{-1/2}\left(\frac{H}{2\pi}\right)^{2}\mathrm{e}^{4\left|\alpha\vartheta\right|}\lesssim10^{-3}\;\mathrm{or}\;1,
\end{equation}
 where we also used the fact that $k/\x a\gg\e{\anm}$ for $c<0$.
Using Eq.~\eqref{eq:Flat-Pz-contr} this becomes
\begin{equation}
\left|\alpha\vartheta\right|<10^{2}\;\mathrm{or}\;10^{4}.\label{eq:Flat-th-BH-bound}
\end{equation}

As shown in Subsection~\eqref{sub:EoMs}, the equation of motion
for the longitudinal mode of the vector field perturbation is not
affected by the axial term. Thus the results of Ref.~\cite{varkin}
for this mode can also be used directly in our case. It was found
that the spectrum of the longitudinal mode is
\begin{equation}
\mathcal{P}_{\|}=\left(\frac{3H}{\e M}\right)^{2}\left(\frac{H}{2\pi}\right)^{2},\label{eq:Flat-Plong}
\end{equation}
 where $M$ is an effective mass of the vector field defined in Eq.~\eqref{eq:M-def}
and the index `$\re$' indicates that it is evaluated at the end of
inflation. In these works it was also found that the spectrum of transverse
modes are unaffected by the mass term if the field is light.

The predominantly parity violating perturbation will be generated
if $\mathcal{P}_{w_{-}}>\mathcal{P}_{\|}$. Comparing Eqs.~\eqref{eq:Flat-P-wm}
and \eqref{eq:Flat-Plong} we find the condition for this to be 
\begin{equation}
\left|\alpha\vartheta\right|^{-3}\mathrm{e}^{4\left|\alpha\vartheta\right|}>\left(\frac{3H}{\e M}\right)^{2}.
\end{equation}
 Taking into account the bound in Eq.~\eqref{eq:Flat-th-BH-bound}
one sees that the above condition can be easily satisfied even for extremely small values of $\e M$.

Results for the power spectrum given in Eqs.~\eqref{eq:Flat-P-wp}
and \eqref{eq:Flat-P-wm} are valid if the vector field is light during
inflation. If this is not the case, and the field becomes heavy before
the end of inflation, it starts oscillating and the amplitude of the
spectrum decreases by $\frac{1}{2}\left(3H/M\right)^{2}$ \cite{varkin}.
As can be seen from the definition of the effective mass $M$ in Eq.~\eqref{eq:M-def}
it can grow only if $\alpha=-4$. In view of this we can rewrite Eqs.~\eqref{eq:Flat-P-wp}
and \eqref{eq:Flat-P-wm} as

\begin{eqnarray}
\mathcal{P}_{w_{+}} & = & \frac{1}{16\pi}\left|\vartheta\right|^{-3}\left(\frac{H}{2\pi}\right)^{2}\frac{1}{2}\mathrm{min}\left\{ 1,\frac{3H}{\e M}\right\} ^{2};\\
\mathcal{P}_{w_{-}} & = & \frac{1}{16\pi}\left|\vartheta\right|^{-3}\left(\frac{H}{2\pi}\right)^{2}\frac{\mathrm{e}^{16\left|\vartheta\right|}}{4}\mathrm{min}\left\{ 1,\frac{3H}{\e M}\right\} ^{2}.
\end{eqnarray}

\subsection{The Vector Curvaton}

So far we have discussed the perturbation of the vector field. One
way a vector field can generate or contribute to the primordial curvature
perturbation $\zeta$ is via the vector curvaton scenario \cite{vecurv}.
In this scenario it is assumed that the vector field is light during
inflation, at least while the cosmological scales exit the horizon.
To avoid excessive anisotropic expansion of the Universe, the energy
density of the light vector field has to be subdominant. After inflation,
when the vector field becomes heavy, it oscillates with a very high
frequency and behaves as preasureless, isotropic matter. The energy
density of such matter decays slower than the radiation, and thus
the vector field can dominate and generate the total of $\zeta$ or
nearly dominate and generate a contribution to $\zeta$
\begin{equation}
\zeta=\left(1-\hat{\Omega}_{W}\right)\zeta_{\mathrm{rad}}+\hat{\Omega}_{W}\zeta_{W},\label{eq:vCurv-z-curv-egn}
\end{equation}
where $\hat{\Omega}_{W}$ is defined in Eq.~\eqref{eq:hOW-def}
and $\zeta_{W}$ is the vector field contribution to the curvature
perturbation $\zeta$. If the vector field perturbation spectrum is
statistically anisotropic, so is the spectrum of $\zeta$ 
\cite{stanis,fnlanis}.

In Ref.~\cite{varkin}
a vector curvaton scenario was studied with the Lagrangian in 
Eq.~\eqref{eq:Flat-L}
and $\vartheta=0$. Bounds derived in these references also apply
to the current scenario as the axial term with $\vartheta\ne0$ does
not contribute to the homogeneous values of fields and, therefore, to the
homogeneous value of the energy density. This is because for the homogenised
vector field, $F_{ij}=0$, where $i,j$ denote spatial components.
The axial term is proportional to 
$\epsilon^{\mu\nu\rho\sigma}F_{\mu\nu}F_{\rho\sigma}$.
This means that, it cannot include any term featuring simultaneously
two factors of the form $F_{0i}=-F_{i0}=\dot{A}$, which is the only
non-zero component of the field strength tensor. Thus to find the
constraint on the energy scale of inflation we can use the expression
for the Hubble parameter found in 
Ref.~\cite{varkin}

\begin{equation}
\frac{H}{\mpl}\sim\ow^{1/2}\zeta_{W}\mathrm{min}\left\{ 1;\frac{\e M}{H_{*}}\right\} ^{-1/3}\mathrm{min}\left\{ 1;\frac{\hat{m}}{\Gamma}\right\} ^{1/12}\left(\frac{\mathrm{max}\left\{ \Gamma_{W};H_{\mathrm{dom}}\right\} }{\Gamma}\right)^{1/4}.\label{eq:vCurv-H-expr}
\end{equation}
 In this equation $H$ is the inflationary Hubble parameter, $\ow$
is defined in Eq.~\eqref{eq:Ow-def}, $\Gamma$ and $\Gamma_{W}$
are the inflaton and vector field decay rates and $H_{\mathrm{dom}}$
is the value of the Hubble parameter after inflation, when the oscillating
curvaton dominates over the radiation bath, if it does not decay earlier.

With the axial term the dominant contribution to the vector field perturbation
comes from the $w_{-}$ term, thus $\mathcal{P}_{w_{-}}\gg\mathcal{P}_{w_{+}},\;\mathcal{P}_{\|}$
and $\mathcal{P}_{+}\approx\frac{1}{2}\mathcal{P}_{w_{-}}$. Using
Eq.~\eqref{eq:gz-expr} we find
\begin{equation}
g_{\zeta}\simeq-\frac{1}{2}\frac{N_{W}^{2}\mathcal{P}_{w_{-}}}{\mathcal{P}_{\zeta}^{\mathrm{iso}}}.
\end{equation}
 Also, using the fact that the anisotropic contribution to the power
spectrum of $\zeta$ must be subdominant \cite{9sigma} (cf.~Eq.~(\ref{grange})),
to the first order we can write
\begin{equation}
\frac{\zeta^{2}}{\delta W^{2}}\simeq\frac{\mathcal{P}_{\zeta}^{\mathrm{iso}}}{\mathcal{P}_{w_{-}}},
\end{equation}
 where we also used the fact that the dominant contribution to the
vector field perturbation is from the $w_{-}$ mode. Combining the
above two equations we find
\begin{equation}
N_{W}^{2}\delta W^{2}\simeq-2g_{\zeta}\zeta^{2}.\label{eq:vCurv-aux-eqn1}
\end{equation}
 Comparing the $\delta N$ formula in Eq.~\eqref{eq:dN-formula}
and the equation for $\zeta$ in the curvaton scenario in Eq.~\eqref{eq:vCurv-z-curv-egn}
we can also write
\begin{equation}
N_{W}\delta W=\hat{\Omega}_{W}\zeta_{W}.
\end{equation}
 Inserting this into Eq.~\eqref{eq:vCurv-aux-eqn1} we get
\begin{equation}
\zeta\sim\frac{\ow\zeta_{W}}{\sqrt{-g_{\zeta}}},\label{eq:vCurv-z-zW-relation}
\end{equation}
 which can be used in Eq.~\eqref{eq:vCurv-H-expr} to find the lower
bound of the inflationary Hubble parameter. In 
Ref.~\cite{varkin}
it was shown that the lowest decay rate of the vector field is through
the gravitational decay, which gives $\mathrm{max}\left\{ \Gamma_{W};H_{\mathrm{dom}}\right\} \geq\e M^{3}/\mpl^{2}$.
Using this and Eq.~\eqref{eq:vCurv-z-zW-relation} we find the bound
on the inflationary Hubble parameter $H$
\begin{equation}
\frac{H}{\mpl}>\left(-\frac{g_{\zeta}\zeta^{2}}{\Omega_{W}}\right)^{1/2}\left(\frac{\e M}{\mpl}\right)^{3/4}\left(\frac{\Gamma}{\mpl}\right)^{-1/4}\mathrm{min}\left\{ 1;\frac{\e M}{H_{*}}\right\} ^{-1/3}\mathrm{min}\left\{ 1;\frac{\hat{m}}{\Gamma}\right\} ^{1/12}.\label{eq:vCurv-H-bound-gen}
\end{equation}
 As we can see from this inequality the bound is maximised if the
Universe undergoes prompt reheating, that is $\Gamma\rightarrow H_{*}$.
Also note, that the bound on $H_{*}$ is maximised for the smallest
value of the effective mass of the vector field $\e M$. Requiring
that the vector field decays before the Big Bang Nucleosynthesis (BBN)
gives $\e M\gtrsim10^{4}\,\mathrm{GeV}$ \cite{varkin}.
Inserting these limits into \eqref{eq:vCurv-H-bound-gen} and using
the fact that $\Omega_{W}<1$, the bound on $H_{*}$ and inflationary
energy scale $V_{*}^{1/4}$ becomes
\begin{equation}
H>\sqrt{-g_{\zeta}}\,10^{6}\,\mathrm{GeV}\quad\Leftrightarrow\quad V_{\mathrm{inf}}^{1/4}>\left(-g_{\zeta}\right)^{1/4}\,10^{12}\,\mathrm{GeV}.\label{eq:vCurv-H-V-bounds}
\end{equation}

Also from the results in Ref.~\cite{varkin} the
constraint on the mass of the vector field is
\begin{equation}
10\,\mathrm{TeV}\lesssim\e M\lesssim10^{6}H,\label{eq:vCurv-Me-bound}
\end{equation}
 where the bound in Eq.~\eqref{eq:vCurv-H-V-bounds} is saturated
with the lower bound in Eq.~\eqref{eq:vCurv-Me-bound}.

From the above two constraints it is clear that there is an ample
parameter space for this scenario to be realised.

\subsection{The Parity Violating Non-Gaussianity}

As we saw in Section~\ref{sec:anisotropic-fNl} to calculate $\fnl$
it is convenient to use parameters $p$ and $q$ defined in Eq.~\eqref{eq:pq-def}.
Because $\mathcal{P}_{w_{-}}$ is exponentially larger than both $\mathcal{P}_{w_{+}}$
and $\mathcal{P}_{\|}$, $p$ and $q$ are equal to

\begin{equation}
p\approx-1\;\mathrm{and}\;\left|q\right|\approx1.\label{eq:Flat-pq-vals}
\end{equation}
The sign of $q$ is determined by the sign of $\alpha\vartheta$.
Following the definitions of $w_{+}$ and $w_{-}$ after Eq.~\eqref{eq:EoM-wpm}
we find that $q\approx-1$ if $\alpha\vartheta>0$ and $q\approx+1$
otherwise.

The value of $\left|p\right|\approx1$ violates observational bounds
on the anisotropy in the spectrum. Thus, as discussed in Section~\eqref{sec:anisotropic-fNl},
the vector field contribution to $\zeta$ must be subdominant, i.e.
$\xi<1$, which gives
\begin{equation}
g_{\zeta}\approx-\xi\quad\mathrm{and}\quad\hat{\Omega}_{W}\approx\frac{3}{4}\Omega_{W}.
\end{equation}

Putting the above values of $p$, $q$, $\xi$ and $\hat{\Omega}_{W}$
in Eqs.~\eqref{eq:fNL-equil}-\eqref{eq:fNL-flat} we find 

\begin{eqnarray}
\frac{6}{5}\fnle & = & 3\frac{g_{\zeta}^{2}}{\Omega_{W}}\left(1-\frac{3}{4}W_{\perp}^{2}\right),\label{eq:fnle-fin}\\
\frac{6}{5}\fnls & = & 2\frac{g_{\zeta}^{2}}{\Omega_{W}}\left(1-W_{\perp}^{2}-i~\mathrm{sgn}(\alpha\vartheta)W_{\perp}\sqrt{1-W_{\perp}^{2}}\sin\omega\right),\\
\frac{6}{5}\fnlf & = & \frac{4}{5}\frac{g_{\zeta}^{2}}{\Omega_{W}}\left(1-\cos^{2}\varphi W_{\perp}^{2}\right),\label{eq:fnlf-fin}
\end{eqnarray}
where $\mathrm{sgn}(\alpha\vartheta)$ is the sign of $\alpha\vartheta$. 
As $\mathcal{P}_{w_{-}}\gg\mathcal{P}_{\|}$, the above result is valid for both: 
the light vector field and the one which becomes heavy
at the end of inflation. 

The shape of $\fnl$ given in Eqs.~\eqref{eq:fnle-fin}-\eqref{eq:fnlf-fin} provides a smoking-gun signature for this model. It is easy to see that the maximum values of $\fnl$ in different configurations are related as
\begin{equation}
\left.\frac13 \fnle\right|_{\rm max} =\left.\mathrm{Re}\left[\frac12\fnls\right]\right|_{\rm max} = \left.\frac54\fnlf\right|_{\rm max},
\end{equation}
while only $\fnle$ has a non-vanishing minimum value of $3/4$.

\section{Concrete examples}\label{ce}

In this section we consider some specific models motivated by particle
physics, which can provide a flat vector field perturbation spectrum
with the use of the axial term $\propto F\tilde{F}$. But before going into the 
models it is important to point out again that the axial term does not affect
the dynamics of the homogenised fields. Thus, we will ignore this
term when discussing the dynamics of the homogenised scalar and vector
fields.

\subsection{String inspired model}

In string theory the gauge kinetic function and the coupling of the axial term 
are of the following form: $f={\rm Re}{\cal F}$ and $h={\rm Im}{\cal F}$,
where ${\cal F}$ is some complex holomorphic function of the moduli fields. 
Thus, the gauge field content of the model is 
\begin{equation}
{\cal L}=-\frac{1}{4}({\rm Re}{\cal F})F_{\mu\nu}F^{\mu\nu}+\frac{1}{2}m^{2}A_{\mu}A^{\mu}+C\,({\rm Im}{\cal F})F_{\mu\nu}\tilde{F}^{\mu\nu}.
\end{equation}
 The simplest form for ${\cal F}$ is ${\cal F}\propto e^{-bT}$,
for large values of the modulus $T$ in Planck mass units ($b=\,$constant).
Writing $T$ in terms of the real fields $\phi$ and $\sigma$ as
$T=\phi+i\sigma$ and reinstating the Planck mass we find 
\begin{equation}
\left.\begin{array}{l}
f=e^{-b\phi/\mpl}\cos(b\sigma/\mpl)\\
\\h=Ce^{-b\phi/\mpl}\sin(b\sigma/\mpl)\end{array}\right\} 
\Rightarrow\frac{h}{f}=\vartheta=C\tan(b\sigma/\mpl),\label{hf}
\end{equation}
 where $\vartheta$ was defined in Eq.~\eqref{eq:Flat-theta-def}.
If the axion $\sigma$ remains frozen during inflation with $\sigma\simeq\mathrm{constant}$
then we have $\vartheta=\mathrm{constant}$, i.e. $f\propto h\propto a^{\alpha}$
and $c=-\frac{1}{2}$. This is a very realistic possibility because
the axion mass is protected by the approximate U(1) symmetry and can
be very small, much smaller than $H$ during inflation. Note that
we have not clarified whether $\phi$ and $\sigma$ are canonically
normalised fields. Indeed, if the K\"{a}hler potential has a non-trivial
dependence on these fields they will not be canonically normalised.
However, this does not change the result above, that is $c=-\frac{1}{2}$.

In the opposite limit of small $T$ the dependence of ${\cal F}$
on the modulus is logarithmic. For toroidal compactifications the
K\"{a}hler potential is of the form \mbox{$K=-3\ln(T+\bar{T})$} in
Planck units, where $T$ is a complex structure modulus here. This
means that \mbox{$K=-3\mpl^{2}\ln(2\phi/\mpl)$}. Hence, the kinetic
term of the $\phi$ field is non-canonical and it reads \[
{\cal L}_{{\rm kin}}=K_{\phi\phi}\partial_{\mu}\phi\partial^{\mu}\phi=3g^{\mu\nu}\left(\frac{\mpl}{\phi}\right)^{2}\partial_{\mu}\phi\partial_{\nu}\phi\,.\]
 In view of this we can define the canonically normalised scalar field
$\Phi$ as \mbox{$\ln(\phi/\mpl)=\frac{1}{\sqrt{6}}\Phi/\mpl$}.
Thus, if we ignore the axion, the gauge kinetic function is $f\propto{\rm Re}{\cal F}\propto(\ln\phi)^{n}\propto\Phi^{n}$,
i.e. it has a power-law dependence on the canonically normalised field
$\Phi$.

In summary, provided we ignore the axion (presumed frozen), string
theory suggests that $h/f=\mathrm{constant}$. The functional dependence
of $f$ and $h$ on the varying modulus (which could be the inflaton)
depends on the (unknown) compactification scheme but both exponential
and power-law dependence is reasonable.

The condition $k/\x a\gg H$ is ensured by the lower bound in Eq.~\eqref{eq:Flat-BD-bound},
i.e. $\vartheta\gg1/4$ with $\alpha=-4$. From Eq.~\eqref{hf},
assuming $C\sim1$, we see that this condition translates into the
lower bound for $\sigma$ 
\begin{equation}
\frac{\sigma}{\mpl}\gg10^{-1}/b\,.\label{eq:s-lower}
\end{equation}
 Thus, assuming $b\sim1$, this condition is satisfied provided the
value of the (frozen) axion is comparable to the Planck scale, but
this can be relaxed if $C\gg1$ (see for example Ref.~\cite{sorbo}).
The upper bound on $\vartheta$,
which limits the overproduction of primordial black holes and is given
in Eq.~\eqref{eq:Flat-th-BH-bound}, translates into 
\begin{equation}
C\tan\left(\frac{b\sigma}{\mpl}\right)<10^{2}\:\mathrm{or}\:10^{4}.\label{eq:s-upper}
\end{equation}
 As was mentioned before, this also ensures that $\e{\anm}\ll H$
for cosmological scales, as the latter is a weaker constraint.
If the two bounds in Eqs.~\eqref{eq:s-lower} and \eqref{eq:s-upper}
are satisfied, it is possible in this set up to generate parity violating
$\zeta$ with a flat power spectrum, as discussed in Section~\ref{sub:Scale-Invariant}.

\subsection{The orthogonal axion}

This model is of the form 
\begin{eqnarray}
{\cal L} & = & D_{\mu}\Phi(D^{\mu}\Phi)^{*}-V(\Phi)-\frac{1}{4}fF_{\mu\nu}F^{\mu\nu}+\hat{c}e^{2}\hat{\theta}F_{\mu\nu}\tilde{F}^{\mu\nu}\nonumber \\
 & = & \frac{1}{2}\partial_{\mu}\phi\partial^{\mu}\phi+\frac{1}{2}\partial_{\mu}\sigma\partial^{\mu}\sigma-V(\phi)-V(\sigma)-\nonumber \\
 &  & -\frac{1}{4}fF_{\mu\nu}F^{\mu\nu}+\frac{1}{2}e^{2}\phi^{2}A_{\mu}A^{\mu}+\hat{c}e^{2}\frac{\sigma}{\phi_{0}}F_{\mu\nu}\tilde{F}^{\mu\nu},\label{eq:Orth-L}\end{eqnarray}
 where $D_{\mu}\equiv\partial_{\mu}+ieA_{\mu}$ and $f$ is modulated
by some inflaton field but $\phi$ or $\sigma$ are not it. Comparing
the above Lagrangian with Eq.~\eqref{eq:Lagrangian} we see that

\begin{equation}
h=-4\hat{c}e^{2}\frac{\sigma}{\phi_{0}}
\end{equation}
 in this model. To have some time-dependence for the axion $\sigma\equiv\hat{\theta}\phi_{0}$
we need its mass to be comparable to the Hubble parameter. Thus, we
consider modular inflation with \mbox{$H\sim\,$TeV} and we assume
that $\sigma$ is an axion field, orthogonal to the QCD axion in supersymmetric
realisations of the Peccei-Quinn symmetry, which employ a non-renormalisable
superpotential for the Peccei-Quinn fields \cite{chun,laza}. 
This construction also solves the $\mu$-problem of supersymmetry.

The superpotential is of the form 
\begin{equation}
W=\frac{\kappa}{n+3}\frac{\Phi^{n+3}}{\mpl^{n}}\,,
\end{equation}
 where \mbox{$\Phi=(\phi/\sqrt{2})\exp(i\hat{\theta}/\sqrt{2})$},
with \mbox{$\hat{\theta}\equiv\sigma/\phi_{0}$} with $\phi_{0}$
being the Peccei-Quinn breaking scale given by 
\begin{equation}
\phi_{0}=2\left(\frac{\mpl^{n}m_{\phi}}{\sqrt{2}\kappa}\right)^{1/(n+1)},\label{phi0}
\end{equation}
 where \mbox{$m_{\phi}\sim\,$TeV} is the tachyonic soft mass of
the radial field $\phi$, which breaks the Peccei-Quinn symmetry.
The above is obtained by minimising 
\begin{equation}
V(\phi)=V_{0}-\frac{1}{2}m_{\phi}^{2}\phi^{2}+\frac{\kappa^{2}}{2^{n+2}}\frac{\phi^{2(n+2)}}{\mpl^{2n}}\,,
\end{equation}
 where $V_{0}$ is some density scale. The orthogonal axion potential
is 
\begin{equation}
V(\sigma)=(\phi m_{\sigma})^{2}[1-\cos(\sigma/\phi)]\,.\label{axionpot}
\end{equation}
 The above potential is due to the soft A-term in the scalar
potential,%
\footnote{\mbox{$\delta V_{{\rm A-term}}={\cal A}(W+W^{*})$} for a monomial
superpotential.%
} which gives the following value for the mass of the orthogonal axion:

\begin{equation}
m_{\sigma}^{2}=\kappa{\cal A}\frac{(\phi/\sqrt{2})^{n+1}}{\mpl^{n}}\,,
\end{equation}
 with \mbox{${\cal A}\sim\,$TeV} being the coefficient of the A-term.
If the radial field assumes its vacuum value \mbox{$\phi=\phi_{0}$}
(i.e. it is not rolling down the radial direction) then the axion
mass becomes 
\begin{equation}
m_{\sigma}^{2}=2^{(n-2)/2}{\cal A}m_{\phi}\;.\label{ms}
\end{equation}

We assume that, during inflation, the axion is very close to the origin
and 
rolls down near a local minimum 
approaching the origin. 
Then we can write 
\begin{equation}
\frac{\partial V(\sigma)}{\partial\sigma}=m_{\sigma}^{2}\phi_{0}\sin(\sigma/\phi_{0})\simeq m_{\sigma}^{2}\sigma\,.
\end{equation}
 Using the above, the axion's equation of motion is 
\begin{equation}
\ddot{\sigma}+3H\dot{\sigma}+m_{\sigma}^{2}\sigma\simeq 0\,,
\end{equation}
 whose growing mode solution is 
\begin{equation}
\sigma\propto\exp\left\{ -\frac{3}{2}\left[1-\sqrt{1-\left(\frac{2}{3}\frac{m_{\sigma}}{H}\right)^{2}}\right]Ht\right\} \propto a^{-\frac{3}{2}\left[1-\sqrt{1-\left(\frac{2}{3}\frac{m_{\sigma}}{H}\right)^{2}}\right]}.\label{eq:Orth-s-sol}
\end{equation}
 Thus, if we denote 
\begin{equation}
h\propto a^{\gamma}
\end{equation}
 from Eq.~\eqref{eq:Orth-s-sol} we find 
\begin{equation}
\gamma=-\frac{3}{2}\left[1-\sqrt{1-\left(\frac{2}{3}\frac{m_{\sigma}}{H}\right)^{2}}\right]\;\Rightarrow\; c=\frac{3}{4}\left[1+\sqrt{1-\left(\frac{2}{3}\frac{m_{\sigma}}{H}\right)^{2}}\right]>0\,,\label{gamma}
\end{equation}
 where we also considered Eqs.~(\ref{eq:N-def}) and (\ref{eq:N-ansatz})
which suggest \mbox{$2c=\gamma+3$}. If \mbox{$m_{\sigma}\ll H$}
then \mbox{$\gamma\approx\frac{1}{3}(\frac{m_{\sigma}}{H})^{2}\ll1\Rightarrow c\simeq3/2$},
i.e. $\sigma$ slow-rolls (and \mbox{$h\propto\ln a$}). In fact,
$\sigma$ is practically frozen, in which case there is no parity
violation. For this reason we do not consider that the radial field
is still rolling with \mbox{$\phi\ll\phi_{0}$}, neither do we assume
that the tachyonic mass of the radial field is suppressed by some
supergravity correction so that \mbox{$(m_{\phi}^{2})_{{\rm eff}}=
m_{\phi}^{2}-\tilde{c}H^{2}\ll m_{\phi}^{2}\sim1\,$TeV}, as in 
Ref.~\cite{laza}.
Both these possibilities would result in \mbox{$m_{\sigma}\ll H\sim1\,$TeV}.
Therefore, we should assume that the Peccei-Quinn symmetry is fully
broken with \mbox{$\phi=\phi_{0}$} and \mbox{$m_{\phi}\sim1\,$TeV},
such that \mbox{$m_{\sigma}\sim\sqrt{{\cal A}m_{\phi}}\sim H\sim1\,$TeV},
according to Eq.~\eqref{ms}.

The above imply that the mass of the physical vector field is 
\begin{equation}
M\equiv\frac{m_{A}}{\sqrt{f}}=\frac{e\phi_{0}}{\sqrt{f}}\propto a^{2}.\label{MA0}
\end{equation}
 Thus, the longitudinal component of the vector field exists and,
if it undergoes particle production, it will not obtain a scale-invariant
spectrum. We need to check whether the longitudinal component spoils
the model.

From Eqs.~\eqref{phi0} and \eqref{MA0} we find 
\begin{equation}
\frac{M}{H}\sim e\left(\frac{\mpl}{m_{3/2}}\right)^{n/(n+1)}e^{-2N}\sim e\times10^{15n/(n+1)}e^{-2N},
\end{equation}
 where \mbox{$m_{3/2}\sim1\,$TeV} stands for the weak scale (gravitino
mass%
\footnote{for gravity mediated supersymmetry breaking.%
}) and we considered that, at the end of inflation, \mbox{$f\rightarrow1$}.
Firstly, we need to verify that the field is light when the cosmological
scales exit the horizon. The vector field becomes heavy (and begins
oscillating) when \mbox{$M\sim H$}. The earliest time for this
to happen can be found by taking \mbox{$e=1$} and \mbox{$n\rightarrow\infty$}
in the above. We obtain \mbox{$N_{{\rm osc}}^{{\rm max}}\simeq\frac{15}{2}\ln10\simeq17$},
which corresponds to much later times than the exit of the cosmological
scales.%
\footnote{\mbox{$N_{*}\simeq50$} for prompt reheating with \mbox{$V_{{\rm \textrm{inf}}}^{1/4}\sim\sqrt{\mpl m_{3/2}}$}.%
}

Secondly, we need to ascertain that the vector field, after it begins
oscillating, survives until the end of inflation at least, so that
it can have some hope to affect the curvature perturbation in the
Universe. The decay rate of the vector field is \mbox{$\Gamma_{W}=\frac{e^{2}}{8\pi}M\propto a^{2}$}.
Requiring that \mbox{$\Gamma_{W}\lsim H$} at the end of inflation
produces the constraint 
\begin{equation}
e\lsim3\times10^{-5n/(n+1)}\sim\left\{ \begin{array}{ll}
0.01 & {\rm for}\; n=1\\
\\10^{-5} & {\rm for}\; n\rightarrow\infty\,.\end{array}\right.
\end{equation}
 This is a tight constraint for the gauge coupling and excludes large
values of $n$. Using the above, we can estimate how close to the
end of inflation the oscillations of the vector field begin. Indeed,
it is straightforward to find 
\begin{equation}
N_{{\rm osc}}\lsim\frac{10n}{n+1}\ln10\sim\left\{ \begin{array}{ll}
6 & {\rm for}\; n=1\\
\\11 & {\rm for}\; n\rightarrow\infty\,.\end{array}\right.
\end{equation}
 Thus, in all cases the field becomes heavy and oscillates a few e-folds
before the end of inflation, but it is always light when the cosmological
scales exit the horizon.

Now, let us consider the particle production process for the longitudinal
component. In Ref.~\cite{varkin} it was found 
\begin{equation}
{\cal P}_{\|}=\frac{16\pi}{\sin^{2}(\pi\hat{\nu})[\Gamma(1-\hat{\nu})]^{2}}\left(\frac{H}{2\pi}\right)^{2}\left(\frac{H}{M}\right)^{2}\left(\frac{k}{2aH}\right)^{5-2\hat{\nu}},
\end{equation}
 where 
\begin{equation}
\hat{\nu}=\frac{1}{2}\sqrt{9+2(\alpha+1)(2-\alpha+2\beta)+(2-\alpha+2\beta)^{2}}\,,
\end{equation}
 with \mbox{$f\propto a^{\alpha}$} and \mbox{$m_{A}\propto a^{\beta}$}.
Using that, in this case, \mbox{$\alpha=-4$} and \mbox{$\beta=0$}
we find \mbox{$\hat{\nu}=\frac{3}{2}$}, which gives 
\begin{equation}
{\cal P}_{\|}=\left(\frac{H}{2\pi}\right)^{2}\left(\frac{H}{M}\right)^{2}\left(\frac{k}{aH}\right)^{2}\Rightarrow\sqrt{{\cal P}_{\|}}=\frac{H}{2\pi}\frac{k}{aM}\propto a^{-3}.
\end{equation}
 As we have discussed, at the end of inflation, \mbox{$M\gg H$}.
Thus, since we are considering superhorizon scales with \mbox{$k\ll aH$},
we see that, at the end of inflation \mbox{${\cal P}_{\|}\ll{\cal P_{\perp}}=(H/2\pi)^{2}$}.
Hence, even though \mbox{${\cal P}_{\|}\propto k^{2}$}, the scale
invariance of the curvature perturbation is not spoilt because the
contribution from the longitudinal component of the vector field is
negligible compared with the transverse components, which are (approximately)
scale-invariant.

In this model flat perturbation spectra for transverse modes will
be produced if $k/\x a\ll H$ and $\e{\anm}\ll H$ (c.f. subsection~\ref{sub:Case-I}).
One can see from the definition of $\anm$ in Eq.~\eqref{eq:N-def}
that $\left(k/\x aH\right)=\x{\left.\left(\left|\dot{h}\right|/f\right)\right|}=\x{\left.\gamma(h/f)\right|}$.
Now, from Eq.~\eqref{gamma} with \mbox{$m_{\sigma}\sim H\sim1\,$TeV},
we get \mbox{$\gamma={\cal O}(1)$}. We also find \mbox{$\x h=\hat{c}e^{2}(\x{\sigma}/\phi_{0})\ll1$},
since \mbox{$\hat{c}={\cal O}(1)$}, \mbox{$e\ll1$} and \mbox{$0<\sigma_{{\rm x}}\ll\pi\phi_{0}$}.
Finally, because \mbox{$\x f=e^{4\x N}\gg1$}, the physical momentum
at $k/\x a=\x{\anm}$ is \mbox{$k/\x a\ll H$} as desired ($\x N$
here denotes the remaining inflationary e-folds when \mbox{$a=\x a$}).
Similarly, $\e{\anm}$ is also smaller than $H$. Since at the end
of inflation, on superhorizon scales $k\ll\e{\left(aH\right)}$ and
$\e{\sigma}\lsim\phi_{0}$ we find $\left(\e{\anm}/H\right)^{2}=\e{\left(\frac{k}{aH}\right)}\left(\frac{\e{\sigma}}{\phi_{0}}\right)\ll1$.
Thus, we see that we do obtain scale-invariant spectra for the perturbations, 
which means that this model can be used to generate statistical anisotropy in
$\zeta$. However, as discussed in the end of Sec.~\ref{sub:Case-I}, in this
case, parity violating signatures are probably undetectable both in
the power spectrum and in the bispectrum, whatever the value of $c$.

\section{Conclusions}\label{conc}

In conclusion, we have demonstrated that it is, in general, possible to employ
an axial coupling for a vector field in order to generate scale invariant 
statistical anisotropy. There are two possibilities for this. 
In the first possibility, the only requirement is that, after horizon exit,
the contribution of the axial term dominates the momentum term in the equation 
of motion for the transverse vector field components (the longitudinal one, if 
it exists, is not affected by the presence of the axial term) but still remains
smaller than the Hubble scale until the end of inflation. In this case, both 
transverse components obtain a scale invariant, superhorizon spectrum of 
perturbations of magnitude $H/2\pi$. If the field is massless then there is no 
longitudinal component (it is decoupled from the theory) so
the anisotropy in the particle production is 100\%, which means that the vector
field has to contribute subdominantly to the curvature perturbation $\zeta$,
as discussed in Ref.~\cite{stanis}. If there is non-zero mass, the longitudinal
component is physical and has to be taken into account, in the way described in
Ref.~\cite{varkin}. The vector field can play the role of vector curvaton and
produce statistical anisotropy in the spectrum and bispectrum of $\zeta$ as 
in Ref.~\cite{varkin}. It can also contribute to $\zeta$ via another 
mechanism, e.g. the end of inflation mechanism \cite{yokosoda,stanis}. We have
presented one example based on particle physics, where the axial coupling 
involves the so-called orthogonal axion, which is orthogonal to the QCD axion 
in supersymmetric realisations of the Peccei-Quinn symmetry \cite{chun,laza}.
In this case, the axial contribution to the equations of motion is growing 
during inflation, while there is a longitudinal component which obtains a 
scale-dependent spectrum, that is subdominant to the scale invariant spectrum 
of the transverse components. However, this possibility does not produce any 
parity violating signatures in the bispectrum.

In contrast, the second possibility can indeed generate a parity violating 
signature. This possibility corresponds to a decreasing contribution of the 
axial coupling to the equation of motion of the transverse vector field mode 
functions. This contribution has to take over the momentum term before horizon 
exit. The scenario can be realised only when the couplings $f$ and $h$ of the 
kinetic and the axial terms respectively are proportional to each other, with 
proportionality constant \mbox{$\vartheta\equiv h/f$}. Then the power spectra 
of the transverse components are
$$
\mathcal{P}_{w_{+}}=\frac{4}{\pi}|\alpha\vartheta|^{-3}(H/2\pi)^2
\qquad{\rm and}\qquad
\mathcal{P}_{w_{-}}=\frac{2}{\pi}|\alpha\vartheta|^{-3}e^{4|\alpha\vartheta|}
(H/2\pi)^2,
$$
i.e. 
\mbox{$\mathcal{P}_{w_{-}}=\frac12 e^{4|\alpha\vartheta|}\mathcal{P}_{w_{+}}$},
where \mbox{$\alpha=-1\pm 3$}. As we have discussed, this possibility can be 
naturally realised in string theory, where $f$ and $h$ are determined by moduli
fields. The parity invariant signature will affect the non-Gaussianity 
by modulating the angular dependence of $f_{\rm NL}$. If the Planck satellite 
does detect statistical anisotropy and anisotropic non-Gaussianity comparison 
between the different configurations (e.g. equilateral, squeezed and flattened)
may well reveal the existence of a parity violating signal and provide evidence
of an axial coupling in the vector field which will be needed to explain the
statistical anisotropy.

\section*{Acknowledgements}
KD thanks E. Kiritsis for stimulating discussions and the University of Crete for
the hospitality. KD is supported by the Lancaster-Manchester-Sheffield Consortium for 
Fundamental Physics under STFC grant ST/J000418/1. MK is supported by CPAN CSD2007-00042
and MICINN (FIS2010-17395) grants.


\end{document}